\documentstyle[11pt]{article}
\input amssymb.sty

\begin{document}
\newcommand{\Si}{\Sigma}
\newcommand{\tr}{{\rm tr}}
\newcommand{\ad}{{\rm ad}}
\newcommand{\Ad}{{\rm Ad}}
\newcommand{\de}{\delta}
\newcommand{\al}{\alpha}
\newcommand{\te}{\theta}
\newcommand{\vth}{\vartheta}
\newcommand{\be}{\beta}
\newcommand{\la}{\lambda}
\newcommand{\La}{\Lambda}
\newcommand{\D}{\Delta}
\newcommand{\ve}{\varepsilon}
\newcommand{\ep}{\epsilon}
\newcommand{\vf}{\varphi}
\newcommand{\G}{\Gamma}
\newcommand{\ka}{\kappa}
\newcommand{\ip}{\hat{\upsilon}}
\newcommand{\Ip}{\hat{\Upsilon}}
\newcommand{\ga}{\gamma}
\newcommand{\ze}{\zeta}
\newcommand{\si}{\sigma}
\newcommand{\na}{\nabla}
\newcommand{\om}{\omega}
\newcommand{\Om}{\Omega}

\def\we{\wedge}

\def\mC{{\mathbb C}}
\def\mZ{{\mathbb Z}}
\def\mR{{\mathbb R}}
\def\mN{{\mathbb N}}

\def\frak{\mathfrak}
\def\gg{{\frak g}}
\def\gJ{{\frak J}}
\def\gS{{\frak S}}
\def\gL{{\frak L}}
\def\gG{{\frak G}}
\def\gk{{\frak k}}
\def\gK{{\frak K}}
\def\gl{{\frak l}}
\def\gh{{\frak h}}

\def\bfa{{\bf a}}
\def\bfb{{\bf b}}
\def\bfc{{\bf c}}
\def\bfd{{\bf d}}
\def\bfm{{\bf m}}
\def\bfn{{\bf n}}
\def\bfp{{\bf p}}
\def\bfu{{\bf u}}
\def\bfv{{\bf v}}
\def\bft{{\bf t}}
\def\bfx{{\bf x}}
\def\bfg{{\bf g}}
\def\bfM{{\bf M}}
\newcommand{\li}{\lim_{n\rightarrow \infty}}
\newcommand{\mat}[4]{\left(\begin{array}{cc}{#1}&{#2}\\{#3}&{#4}
\end{array}\right)}
\newcommand{\thmat}[9]{\left(
\begin{array}{ccc}{#1}&{#2}&{#3}\\{#4}&{#5}&{#6}\\
{#7}&{#8}&{#9}
\end{array}\right)}
\newcommand{\beq}[1]{\begin{equation}\label{#1}}
\newcommand{\eq}{\end{equation}}
\newcommand{\beqn}[1]{\begin{eqnarray}\label{#1}}
\newcommand{\eqn}{\end{eqnarray}}
\newcommand{\p}{\partial}
\newcommand{\di}{{\rm diag}}
\newcommand{\ti}{\tilde}
\newcommand{\oh}{\frac{1}{2}}

\newcommand{\su}{{\bf su_2}}
\newcommand{\uo}{{\bf u_1}}
\newcommand{\GL}{{\rm GL}(N,{\mC})}
\newcommand{\SLN}{{\rm SL}(N,{\mC})}
\newcommand{\SLt}{{\rm SL}(3,{\mC})}
\def\sln{{\rm sl}(N,\mC)}
\def\gln{{\rm gl}(N,\mC)}
\newcommand{\PSL}{{\rm PSL}_2({\mZ})}
\def\SL2{{\rm SL}(2,\mC)}

\newcommand{\ran}{\rangle}
\newcommand{\lan}{\langle}
\def\f1#1{\frac{1}{#1}}
\def\lb{\lfloor}
\def\rb{\rfloor}
\newcommand{\rar}{\rightarrow}
\newcommand{\upar}{\uparrow}
\newcommand{\sm}{\setminus}
\newcommand{\ms}{\mapsto}
\newcommand{\bp}{\bar{\partial}}
\newcommand{\bz}{\bar{z}}
\newcommand{\bA}{\bar{A}}
\newcommand{\ba}{\bar{a}}
\newcommand{\sect}[1]{\setcounter{equation}{0}\section{#1}}
\renewcommand{\theequation}{\thesection.\arabic{equation}}
\newtheorem{predl}{Proposition}[section]
\newtheorem{defi}{Definition}[section]
\newtheorem{rem}{Remark}[section]
\newtheorem{cor}{Corollary}[section]
\newtheorem{lem}{Lemma}[section]
\newtheorem{theor}{Theorem}[section]

\vspace{0.3in}
\begin{flushright}
 ITEP-TH-15/00\\
\end{flushright}
\vspace{10mm}
\begin{center}
{\Large\bf
Hamiltonian Algebroids and deformations\\
of complex structures on Riemann curves}\\
\today\\
\vspace{5mm}
A.M.Levin\\
{\sf Institute of Oceanology, Moscow, Russia,} \\
{\em e-mail alevin@wave.sio.rssi.ru}\\
M.A.Olshanetsky
\\
{\sf Institute of Theoretical and Experimental Physics, Moscow, Russia,}\\
{\em e-mail olshanet@heron.itep.ru}\\

\vspace{5mm}
\end{center}
\begin{abstract}
Starting with a Lie algebroid ${\cal A}$ over a space $M$ we  lift
its action to the canonical transformations on the affine bundle
${\cal R}$ over the cotangent bundle $T^*M$. Such lifts are classified
by the first cohomology $H^1({\cal A})$. The resulting object is a
Hamiltonian algebroid ${\cal A}^H$ over ${\cal R}$ with the anchor map from
$\G({\cal A}^H)$ to Hamiltonians of canonical transformations.
Hamiltonian algebroids generalize Lie algebras of
canonical transformations. We prove that the BRST operator for ${\cal A}^H$
is cubic in the ghost fields as in the Lie algebra case.
The Poisson sigma model is a natural example of this construction.
Canonical transformations of its phase space define a Hamiltonian algebroid
with the Lie brackets related to the Poisson structure on the target space.
We apply this scheme to analyze the symmetries of generalized deformations
of complex structures on Riemann curves $\Si_{g,n}$ of genus $g$
with $n$ marked points.
We endow the space of local $\GL$-opers with the Adler-Gelfand-Dikii (AGD)
Poisson brackets. Its allows us to define a Hamiltonian algebroid over
the phase space of $W_N$-gravity on $\Si_{g,n}$. The sections of the
algebroid are Volterra operators on $\Si_{g,n}$   with the Lie brackets
coming from the AGD bivector. The symplectic reduction
defines the finite-dimensional moduli space of $W_N$-gravity and in particular
the moduli space of the complex structures $\bp$ on $\Si_{g,n}$ deformed
by the Volterra operators.
\end{abstract}

\vspace{0.15in}
\bigskip
\title
\maketitle
\tableofcontents
\section {Introduction}
\setcounter{equation}{0}
 Lie groups by no means exhaust symmetries in gauge theories.
Their importance is related to the natural geometric structures
defined by a group action in accordance with the Erlanger
program of F.Klein. The first class constraints in Hamiltonian systems
generate the canonical transformations of the phase space which generalize the
Lie group actions \cite{HT}.
There exists a powerful approach to treat such
types of structures. It is the BRST method that is applicable in
Hamiltonian and Lagrangian forms \cite{BFV}. The BRST operator
corresponding to arbitrary first class constraints acquires the most general
form. An intermediate step in this direction is the canonical transformations
generated by the quasigroups \cite{Ba,KM}.
The BRST operator for the quasigroup action has the same form as for
the Lie group case.

Here we consider the quasigroup symmetries
constructed by means of special transformations of the "coordinate
space" $M$. These transformations along with the coordinate space $M$ define
the Lie groupoids, or their infinitesimal version - the Lie algebroids
${\cal A}$ \cite{Ma,We}.
We lift the algebroid action from $M$ to the cotangent bundle $T^*M$,
or, more generally, to the
 principle homogeneous space ${\cal R}$ over
the cotangent bundle $T^*M$. We call this bundle
the Hamiltonian algebroid ${\cal A}^H$ related to the Lie algebroid ${\cal A}$.
The Hamiltonian algebroid is an analog of the Lie algebra of symplectic
vector fields with respect to the canonical symplectic structure on
${\cal R}$ or $T^*M$.
The lifts  from $M$ to ${\cal R}$'s are classified by the first cohomology group $H^1({\cal A})$.
We prove that the BRST operator of ${\cal A}^H$ has the same structure as
for the Lie algebras transformations.

The general example of this construction is the Poisson sigma-model \cite{I,SS}.
The Lie brackets of the Hamiltonian algebroid over the
phase space of the Poisson sigma-model are defined by the Poisson bivector on the target space $M$.

Our main interest lies in topological field
theories,
where the factorization with respect to the canonical gauge transformations
may lead to generalized  deformations of corresponding moduli spaces.
We apply
this scheme to analyze the moduli space of deformations of the
complex structures on Riemann curves
of genus $g$ with $n$  marked points $\Si_{g,n}$
by differential operators of finite order, or equivalently by the
Volterra operators.

To define the deformations
we start with the space $M_N$ of $\SLN$-opers over $\Si_{g,n}$ \cite{Tel,BD},
 and define a Lie algebroid ${\cal A}_N$ over $M_N$. The Lie brackets on the space of sections
$\G({\cal A}_N)$ are derived from the Adler-Gelfand-Dikii brackets for the
local opers over a punctured disk $D^*$ \cite{Ad,GD}. In this way the set of
the local opers serves as the target space of the Poisson sigma-model.
The space $M_2$ of $\SL2$-opers is the space of the projective structures on
$\Si_{g,n}$ and the Lie algebroid ${\cal A}_2$ leads to the Lie algebra of vector
fields on $\Si_{g,n}$. The case $N>2$ is more subtle and we deal with a genuine
Lie algebroids since differential operators of order greater than one do
not form a Lie algebra with respect to the standard commutator.
The AGD brackets define a new commutator on $\G({\cal A}_N)$ that depends on
the projective structure and higher spin fields. In other words, for $N>2$
we deal with the structure functions rather than with the structure constants.

The space $M_N$ of $\SLN$-opers can be considered as a configuration space
of $W_N$-gravity \cite{P,BFK,GLM}. The whole phase space ${\cal R}_N$
of $W_N$-gravity is the
affinization of the cotangent bundle $T^*M_N$. Its sections define the
deformations of the operator $\bp$ by the Volterra
operators.  The canonical transformations of ${\cal R}_N$ are
sections of the Hamiltonian algebroid ${\cal A}_N^H$ over ${\cal R}_N$.
The symplectic quotient of the phase space
is the so-called ${\cal W}_N$-geometry of $\Si_{g,n}$. Roughly speaking, this space is
a combination of the moduli of generalized complex structures
 and the spin $2\,\ldots$,\,spin $N$ fields as the dual variables.
To define the ${\cal W}_N$-geometry
we construct the BRST operator for the Hamiltonian algebroid. As it follows from
the general construction, it has the same structure as in the Lie algebra case.
We consider in detail the simplest nontrivial case $N=3$. It is possible
in this case to describe
explicitly the sections of the algebroid as the second order differential operators,
instead of Volterra operators. This algebroid is
generalization
of the Lie algebra vector fields on $\Si_{g,n}$.
It should be noted that the BRST operator for the $W_3$-algebras was constructed
in \cite{TM}. But here we construct the BRST operator for the different object
- the algebroid symmetries of $W_3$-gravity.  Recently, another BRST description
of $W$-symmetries  was proposed in Ref.\cite{BL}.
We explain our formulae and the origin of the algebroid
by the special gauge procedure of the $\SLN$ Chern-Simons theory using an
approach developed in Ref.\cite{BFK}.

The paper is organized as follows. In the next section we define the general
Hamiltonian algebroids, their cohomolgies and the BRST construction.
We also introduce a special class of Hamiltonian algebroids  related to
Lie algebroids and prove that the BRST has the Lie algebraic form.
In Section 3 we treat the Poisson sigma model and its symmetries as a
Hamiltonian algebroid  related to a Lie algebroid. In Section 4 we consider
two examples of our construction when the algebroids coincide with Lie algebras.
Namely, we analyze the moduli space of flat $\SLN$-bundles
and the moduli of projective structures on $\Si_{g,n}$.
A nontrivial example of this construction is $W_3$-gravity.
It is considered in detail in Section 5.
The general case $W_N$ is analyzed in Section 6.

\section{Hamiltonian  algebroids and  groupoids}
\setcounter{equation}{0}

\subsection{Lie algebroids and groupoids}

We start with a brief description of Lie algebroids and Lie groupoids.
Details of this theory can be found in \cite{KM,Ma,We}.
\begin{defi}
A {\em Lie algebroid} over a differential manifold $M$ is a vector bundle
${\cal A}\rar M$ with
a Lie algebra structure on the space of its sections $\G({\cal A})$
 defined by the Lie brackets $\lb\ve_1,\ve_2\rb,\\
\ve_1,\ve_2\in\G({\cal A})$ and
a bundle map ({\em the anchor}) $\de :{\cal A}\to TM$, satisfying the following
conditions:\\
(i) For any $\ve_1,\ve_2\in\G({\cal A})$
\beq{5.1}
[\de_{\ve_1},\de_{\ve_2}]=\de_{\lb\ve_1\ve_2\rb}\,,
\eq
(ii) For any $\ve_1,\ve_2\in\G({\cal A})$ and $f\in C^\infty(M)$
\beq{5.2}
\lb\ve_1,f\ve_2\rb=f\lb\ve_1,\ve_2\rb + (\de_{\ve_1} f)\ve_2\,.
\eq
\end{defi}
In other words, the anchor  defines a representation of $\G({\cal A})$
in the Lie algebra of vector fields
on $M$. The second condition is the Leibnitz rule with respect to the multiplication of
the sections by smooth functions.

Let $\{e^j(x)\}$ be a basis of local sections $\G ({\cal A})$. Then  the brackets
are defined by the structure functions $f^{jk}_i(x)$ of the algebroid
\beq{5.1b}
\lb e^j,e^k\rb=f^{jk}_i(x)e^i\,,~~x\in M\,.
\eq
Using the Jacobi identity for the anchor action, we find
\beq{5.3}
C^n_{j,k,m}\de_{e_n}=0\,,
\eq
where
\beq{5.4}
C^n_{jkm}=f^{jk}_i(x)f_n^{im}(x)+\de_{e^m}f^{jk}_n(x)+{\rm c.p.}(j,k,m)\,.
\footnote{The sums over repeated indices are understood throughout the paper.}
\eq
Thus, (\ref{5.3}) implies {\em the anomalous Jacobi identity} (AJI)
\beq{5.5}
f^{jk}_i(x)f_n^{im}(x)+\de_{e^m}f^{jk}_n(x)+{\rm c.p.}(j,k,m)=0\,.
\eq

Here are some examples of Lie algebroids.\\
1)If the anchor is trivial, then ${\cal A}$ is just a bundle of Lie algebras.\\
2)Consider an integrable system that has the Lax representation
$$
\p_t{\cal L}=[{\cal L},{\cal M}]\,.
$$
The Lax operator ${\cal L}$ belongs to some subvariety $M$ of an ambient
space ${\cal R}$. In many cases it is
a Lie coalgebra. The second operator ${\cal M}$ defines the tangent vector field
to $M$.
The operators ${\cal M}$ are sections of the Lie algebroid ${\cal A}_M$ over
$M$ with the anchor determining by the Lax equation.
In the similar way the dressing transformations
are sections of the algebroid over $M$ \cite{Wei}.\\
3) Lie algebroids can be constructed from Lie algebras.
Let $G$ be a Lie group that act on a space $S$ and $P$ is subgroup of $G$. Consider the set
of orbits $M=S/P$. For $x\in M$ we have the decomposition the tangent space
\beq{a1}
T_xS=T_xM\oplus T_x{\cal O}_P\,,
\eq
where  ${\cal O}_P $ is the orbit of $P$ containing $x$.
Let $\ep $ be an element of the Lie algebra $\gG$
and Pr$_P$ be the projection on the second term in (\ref{a1}).
 Impose the following condition on the vector field $\de_\ep$
\beq{a2}
{\rm Pr}_P\de_\ep(x)=0
\eq
The subspace of the Lie algebra $\gG$ that satisfy this condition
is the set of the sections of the Lie algebroid ${\cal A}_M(\gG)$.
The anchor is defined by the first term in (\ref{a1}).
The commutators on the sections is the commutators of $\gG$.

In a generic case a Lie algebroid can be integrated to a global object
 - the {\em  Lie groupoid} \cite{Ba,KM,Ma,We}.
\begin{defi}
A Lie groupoid $G$ over a manifold $M$
is a pair of differential manifolds $(G,M)$,  two
differential mappings $l,r~: ~G\to M$ and a partially defined binary operation
(a product)
$(g,h)\mapsto g\cdot h $ satisfying the following conditions:\\
(i) It is defined only when $l(g)=r(h)$. \\
(ii) It is associative: $(g\cdot h)\cdot k=g\cdot (h\cdot k)$
whenever the products are defined.\\
(iii) For any $g\in G$ there exist the left and right identity elements $l_g$ and
$r_g$ such that $l_g \cdot g=g\cdot r_g=g$.\\
(iv) Each $g$ has an inverse $g^{-1}$ such that $g\cdot g^{-1}=l_g$ and
$g^{-1}\cdot g=r_g$.\\
\end{defi}
We denote an element of $g\in G$ by the triple $<< x|g|y>>$, where $x=l(g),~y=r(g)$.
Then  the product $g\cdot h $ is
$$
g\cdot h\rar <<x|g\cdot h|z>>=<<x|g|y>><<y|h|z>>\,.
$$
An orbit of the groupoid in the base $M$ is defined as an equivalence
$x\sim y$ if $x=l(g),~y=r(g)$. The isotropy subgroup $G_x$ for $x\in M$ is
defined as
$$
G_x=\{g\in G~|~l(g)=x=r(g)\}\sim\{<<x|g|x>>\}\,.
$$

The Lie algebroid is a local version of the Lie groupoid. The anchor is determined in
terms of the multiplication law.
Details can be found in \cite{Ba}.

\subsection{Lie algebroid representations and Lie algebroid cohomology}

The definition of algebroids representations is rather evident:
\begin{defi}
A vector bundle representation (VBR) $(\rho, {\cal M})$ of the Lie algebroid
$\cal A$ over $M$
is a vector bundle $\cal M$ over $M$ and a map
 $\rho$ from $ {\cal A}$ to the bundle of differential operators on $\cal M$ of the order
 less or equal to $1$ $\mbox{\it Diff}^{~\le 1}({\cal M}, {\cal M})$,
 such that:\\
(i) the symbol of $\rho(\ve)$ is a scalar equal to the anchor of $\ve$:
$$
{\rm Symb}(\rho(\ve))={\rm Id}_{\cal M}\de_\ve\,,
$$
(ii) for any $\ve_1,\ve_2\in\G({\cal A})$
\beq{rep}
[\rho({\ve_1}),\rho({\ve_2})]=\rho({\lb\ve_1\ve_2\rb})\,,
\eq
 where  the l.h.s. denotes
the commutator of differential operators.
\end{defi}

For example, the trivial bundle is a VBR representation (the map $\rho$
is the anchor map $\delta$),

Consider a small disk $U_{\alpha} \subset M$ with local coordinates
$x=(x_1,\ldots,x_a,\ldots)$. Then the anchor can be written as
\beq{5.15i}
\de_{e^j}=b^j_a(x)\frac{\p}{\p x_a}=\lan b^j|\frac{\de}{\de  x}\ran   \,.
\footnote{The brackets $\lan|\ran   $ mean summations over all indices,
 taking a traces,
integrations, etc.}
\eq
Let $w$ be a section of the tangent bundle $TM$. Then the VBR on
 $TM$ takes
the form
\beq{5.15j}
{\rho}_{e^j}w=\lan  b^j|\frac{\de}{\de x}w\ran   -\lan\frac{\de}{\de x}b^j|w\ran   \,.
\eq
Similarly, the VBR the action of $\rho$ on a section $p$ of $T^*M$ is
\beq{5.16i}
\rho_{e^j}p=\frac{\de}{\de x}\lan  p|b^j(x)\ran \,  .
\eq
We drop a more general definition of the sheaf representation.

Now we define  cohomology groups of  algebroids.
First, we consider the case of contractible base $M$.
Let ${\cal A}^*$ be a bundle over $M$ dual to ${\cal A}$.
 Consider the bundle of graded commutative algebras
$\wedge^\bullet{\cal A}^*$.
The space $\G(M, \wedge^\bullet {\cal A}^*)$
is generated by the sections $\eta_k$:
$\lan\eta_j|e^k\ran   =\de_j^k$. It is a graded algebra
$$
\G(M, \wedge^\bullet  {\cal A}^*)=\oplus {\cal A}^*_n\,,~~
{\cal A}^*_n=\{c_n(x)=
\f1{n!}c_{j_1,\ldots,j_n}(x)\eta_{j_1}\ldots\eta_{j_n},~x\in M\}\,.
$$

 Define the Cartan-Eilenberg operator ``dual'' to the  brackets
 $\lb ,\rb$
$$
sc_n(x;e^{1},\ldots,e^{n},e^{n+1})=
(-1)^{i-1}\de_{e^i}c_n(x;e^{1},\ldots,\hat{e}^{i},e^{n})-
$$
\beq{5.6}
-\sum_{j<i}
(-1)^{i+j}c_n(x;\lb e^{i},e^j\rb,\ldots,\hat{e}^{j},\ldots,\hat{e}^{i},\ldots,e^{n}).
\eq
It follows from (\ref{5.1}) and AJI (\ref{5.5}) that $s^2=0$.
Thus,
$s$ determines a complex of bundles ${\cal A}^*\to \wedge^2{\cal A}^*\to \cdots$.

The cohomology groups of this complex are called {\em the cohomology groups of
algebroid with trivial coefficients}.
This complex is  a part of the BRST complex
derived below.

 The action of the coboundary operator $s$ takes the following form
 on the lower cochains:
\beq{5.7}
sc(x;\ve)=\de_\ve c(x)\,,
\eq
\beq{5.8}
sc(x;\ve_1,\ve_2)=\de_{\ve_1}c(x;\ve_2)-\de_{\ve_2}c(x;\ve_1)-
c(x;\lb \ve_1,\ve_2\rb)\,,
\eq
\beq{5.9}
sc(x;\ve_1,\ve_2,\ve_3)=\de_{\ve_1}c(x;\ve_2,\ve_3)-
\de_{\ve_2}c(x;\ve_1,\ve_3)
\eq
$$
+\de_{\ve_3}c(x;\ve_1,\ve_2)
-c(x;\lb \ve_1,\ve_2\rb,\ve_3)
+c(x;\lb \ve_1,\ve_3\rb,\ve_2)
-c(x;\lb \ve_2,\ve_3\rb,\ve_1)\,.
$$

It follows from (\ref{5.7}) that $H^0({\cal A},M)$  is isomorphic to the
invariants in the space $C^\infty (M)$. The next cohomology group
$H^1({\cal A},M)$ is responsible for the shift of the anchor action:
\beq{5.11}
\hat{\de}_\ve f(x)= \de_\ve f(x)+c(x;\ve), ~~sc(x;\ve)=0.
\eq
If $c(x;\ve)$ is a cocycle (see (\ref{5.8})), then this action is
 consistent with the defining anchor property (\ref{5.1}).
 The action (\ref{5.11}) on $\Psi=\exp f(x)$
takes the form
\beq{5.11a}
\hat{\de}_{\ve}\Psi=\left(\de_\ve+c(x;\ve)\right)\Psi(x).
\eq
This formula defines a ``new'' structure of VBR on the {\em trivial } line
bundle.
Let $\ti{M}=M/G$ be the
set of orbits of the groupoid $G$ on its base $M$.
The condition
\beq{5.11b}
\hat{\de}_{\ve}\Psi=0
\eq
defines a linear bundle ${\cal L}(\ti{M})$ over $\ti{M}$.

Two-cocycles $c(x;\ve_1,\ve_2)$ allow to construct
the central extensions of brackets on $\G({\cal A})$
\beq{5.9a}
\lb(\ve_1,k_{1}),(\ve_2,k_{2})\rb_{c.e.}=
(\lb\ve_1,\ve_2\rb,c (x;\ve_1,\ve_2))\,.
\eq
The cocycle condition (\ref{5.9}) means that the new brackets
$\lb~,~\rb_{c.e.}$ satisfies AJI (\ref{5.5}).
The exact cocycles leads to the splitted extensions.

If $M$ is not contractible the definition of cohomology group is more
complicated. We sketch the \^{C}ech version of it.
Choose an acyclic covering $U_\alpha$. Consider the \^{C}ech
complex with coefficients in $\bigwedge{}^{\bullet}({\cal A}^{*})$
corresponding to this covering:
$$\bigoplus \Gamma (U_\alpha ,\bigwedge{}^{\bullet}({\cal
A}^{*}))
\stackrel{d}{\longrightarrow}
\bigoplus\Gamma (U_{\alpha\beta },\bigwedge{}^{\bullet}({\cal
A}^{*}))
\stackrel{d}{\longrightarrow}\cdots$$
The \^{C}ech differential $d$ commutes with  the Cartan-Eilenberg operator
$s$, and cohomology of algebroid are cohomology of normalization of this
bicomplex~:
$$\bigoplus \Gamma (U_\alpha ,{\cal
A}^{*}_0)
\stackrel{d,s}{\longrightarrow}
\bigoplus\Gamma (U_{\alpha\beta },{\cal
A}^{*}_0)
\oplus
\bigoplus \Gamma (U_\alpha ,{\cal
A}^{*}_1)
\stackrel{
\left(\begin{array}{ccc}
d&s&\\
&-d&s
\end{array}\right)
}{\longrightarrow}$$
$$
\bigoplus\Gamma (U_{\alpha\beta\gamma },{\cal
A}^{*}_0)
\oplus
\bigoplus \Gamma (U_{\alpha\beta} ,{\cal
A}^{*}_1)
\oplus
\bigoplus \Gamma (U_\alpha ,{\cal
A}^{*}_2)\longrightarrow \cdots\,.
$$
The cochains
 $c^{i,j}\in\bigoplus_{\alpha_{1}\alpha_{2}\cdots \alpha_j}
\Gamma (U_{\alpha_{1}\alpha_{2}\cdots \alpha_j} ,{\cal
A}^{*}_i)$ are bigraded. The differential maps $c^{i,j}$ to $(-1)^jd\,c^{i,j}+sc^{i,j}$,
 has type $(i,j+1)$ for $(-1)^jd\,c^{i,j}$ and $(i+1,j)$ for $sc^{i,j}$.

Again, the group $H^0({\cal A},M)$  is isomorphic to the
invariants in the whole space $C^\infty (M)$.

Consider the next group $H^{(1)}({\cal A},M)$.
It has  two components\\ $(c_\alpha(x,\ve),c_{\alpha\beta}(x))$.
They are characterized by the following conditions (see (\ref{5.8}))
 $$
c_\alpha(x;\lb \ve_1,\ve_2\rb)=
\de_{\ve_1}c_\alpha(x;\ve_2)-\de_{\ve_2}c_\alpha(x;\ve_1)\,,
$$
\beq{ch1}
  \de_{\ve}c_{\alpha\beta}(x)=-c_\alpha(x;\ve)+c_\beta(x;\ve)\,,
\eq
\beq{ch2}
 c_{\alpha\gamma}(x)=c_{\alpha\beta}(x)+c_{\beta\gamma}(x)\,.
\eq
While the first component  $c_\alpha(x,\ve)$ comes from the algebroid action
on $U_\alpha$ and define the action of the algebroid on the trivial bundle
(\ref{5.11a}), the second
component  determines a line bundle ${\cal L}$ on $M$ by the transition functions
$\exp (c_{\alpha\beta})$. The condition (\ref{ch1}) shows that the actions on
the restriction to $U_{\alpha\beta}$ are compatible.

The continuation of the central extension (\ref{5.9a}) from $U_\al$ on $M$ is defined now
by $H^{(2)}({\cal A},M)$.
There is an obstacle to this continuations in $H^{(3)}({\cal A},M)$.
We do not dwell on this point.

\subsection{Hamiltonian algebroids}

We modify the notion of the Lie algebroids in the following way.
Let ${\cal R}$ be a Poisson manifold.
 Any smooth function $h\in C^\infty({\cal R})$
 gives rise to a vector field
$$
\de_{h} x=\{x,h\}\,.
$$
The space $C^\infty({\cal R})$ has
the structure of a Lie algebra with respect to the Poisson brackets.
In what follows we assume that ${\cal R}$ is a symplectic manifold with
the symplectic form $\om$. In this case the Poisson brackets
$$
\{h_1,h_2\}=-i_{h_1}dh_2\,.
$$
are defined by the internal derivation $i_h$ of
the symplectic form $i_h\om=d h$.

Let ${\cal A}^H$ be a vector bundle over ${\cal R}$ and
assume that the space of sections $\G({\cal A}^H)$ is equipped by
 the Lie brackets $\lb\ve_1,\ve_2\rb$.
\begin{defi}
${\cal A}^H$ is a {\em Hamiltonian  algebroid} over a Poisson
manifold ${\cal R}$ if there exists
a bundle map from ${\cal A}^H$ to the Lie algebra
on $C^\infty({\cal R})$: $\ve\to h_\ve$, (i.e. $f\ve\to fh_\ve$ for
$f\in C^\infty({\cal R})$)
satisfying the following conditions:\\
(i) For any $\ve_1,\ve_2\in\G ({\cal A}^H)$
\beq{5.12}
\{ h_{\ve_1},h_{\ve_2}\}=h_{\lb\ve_1,\ve_2 \rb}\,.
\eq\\
(ii) For any $\ve_1,\ve_2\in\G ({\cal A}^H)$ and $f\in C^\infty({\cal R})$
$$
\lb\ve_1,f\ve_2\rb=f\lb\ve_1,\ve_2\rb+\{h_{\ve_1},f\}\ve_2\,.
$$
\end {defi}
The both conditions are similar to the defining properties of the
Lie algebroids (\ref{5.1}),(\ref{5.2}).
\begin{rem}
In contrast with the Lie algebroids with the anchor $\de_\ve$, that is a
bundle map:
$f\ve\to f\de_\ve$,
 for the Hamiltonian algebroids
one has the map to the first order differential operators with respect to $f$
$$
f\ve\to f\de_{h_\ve}+h_\ve\de_f\,.
$$
\end{rem}

Let $f^{jk}_i$ be structure functions of a Hamiltonian algebroid  and
$$
C_n^{j,k,m}=f^{jk}_i(x)f_n^{im}(x)+\{h_{e^m},f^{jk}_n(x)\}+{\rm c.p.}(j,k,m)\,.
$$
Then the Jacobi identity for the Poisson brackets implies
\beq{5.13a}
C_n^{j,k,m}h_{\ve^n}=0\,.
\eq
This identity is similar to (\ref{5.3}) for Lie algebroids.
But now one can add to $C_n^{j,k,m}$  the term proportional to
$E^{j,k,m}_{[ln]}h_{\ve^l}$ without the breaking (\ref{5.13a})
(here $[,]$ means the antisymmetrization).
Thus, the Jacobi identity for the Poisson algebra of Hamiltonians leads
to following identity for the structure functions
\beq{5.14}
f^{jk}_i(x)f_n^{im}(x)+\{h_{e^m},f^{jk}_n(x)\}+
E^{j,k,m}_{[ln]}h_{\ve^l}+c.p.(j,k,m)=0\,.
\eq
This structure arises in the Hamiltonian systems with the first class constraints
\cite{Ba} and leads to the so-called open algebra of an arbitrary rank (see \cite{HT,BFV}).

The important particular case
\beq{5.15}
f^{jk}_i(x)f_n^{im}(x)+\{H_{\ve^m}f^{jk}_n(x)\}+c.p.(j,k,m)=0
\eq
corresponds to the open algebra of rank one similar to the
Lie algebroid (\ref{5.5}). We will call (\ref{5.15}) a simple anomalous
Jacobi identity (SAJI) preserving the notion AJI for the general form (\ref{5.14}).
In this case the Hamiltonian algebroid can be integrated to the Hamiltonian groupoid.
The later is the Lie groupoid with the canonical action
with respect to the symplectic form on the base ${\cal R}$.

\subsection{Symplectic affine bundles over cotangent bundles}

We shall define Hamiltonian algebroids over cotangent bundles which
are a special class of symplectic manifold. There exist a slightly more general
symplectic manifolds than cotangent bundles, that we include in our
scheme. It is  an affinization over a cotangent bundle we are going to
define. Let $M$ be a vector space and ${\cal R}$ is a set with an action of
$M$ on ${\cal R}$
$$
{\cal R}\times M\rar {\cal R}:(x,v)\to x+v\in {\cal R}\,.
$$
\begin{defi}
The set ${\cal R}$ is an {\sl affinization} over $M$ (a {\em  principle homogeneous
space} over $M$) ${\cal R}/M$ if the action of $M$ on  ${\cal R}$ is transitive
and exact.
\end{defi}
In other words, for any pair $x_1,x_2\in{\cal R}$ there exists $v\in M$ such that
$x_1+v=x_2$, and $x_1+v\neq x_2$ if $v\neq 0$.

This construction is generalized on bundles. Let $E$ be a bundle over $M$ and
$\G(U,E)$ be the linear space of sections in a trivialization of $E$ over some
 disk $U$.
\begin{defi}
{\em An affinization} ${\cal R}/E$ of  $E$
 is a bundle over $M$ with the space of
local sections $\G(U,{\cal R})$ defined as the affinization over $\G(U,E)$.
\end{defi}
Two affinizations ${\cal R}_1/E$ and ${\cal R}_2/E$ are equivalent if there
exists a bundle map compatible with the action of the corresponding vector bundles.
It can be proved that non-equivalent affinizations are classified by $H^1(M,\G(E))$.

Let $E=T^*M$.
Consider a linear bundle ${\cal L}$ over $M$. The space of connections
Conn$_M({\cal L})$ can be identified with the space of sections
${\cal R}/T^*M$. In fact, for any
connection $\nabla_x, ~x\in U\subset M$ one can define another connection
$\nabla_x+\xi,~\xi\in \G(T^*M)$. The affinization ${\cal R}/T^*M$ can be classified by
the first Chern class $c_1({\cal L})$. The trivial bundles correspond to
$T^*M$.

The affinization ${\cal R}/T^*M$ is
the symplectic space with the canonical form $\lan  dp\wedge dx\ran   $. In contrast with $T^*M$
this form is not exact, since $pdx$ is defined only locally.
In the similar way as for $T^*M$,
the space of square integrable sections $L^2(\G({\cal L}))$ plays the role of
the Hilbert space in the prequantization of the affinization ${\cal R}/T^*M$.
For $f\in{\cal R}$ define the Hamiltonian vector field $\al_f$ and the
covariant derivative $\nabla(f)_x=i_{\al_f}\nabla_x $   along $\al_f$.
Then the prequantization of ${\cal R}/T^*M$ is determined by the operators
$$
\rho(f)=\f1{i}\nabla(f)_x+f
$$
acting on the space $L^2(\G({\cal L}))$. In particular,
$\rho(p)=\f1{i}\frac{\de}{\de x}$,
$\rho(x)=x$.

The basic example, though for infinite-dimensional spaces, is the affinizations over
the antiHiggs bundles.
\footnote{We use the antiHiggs bundles instead of the standard Higgs bundles for
reasons, that will become clear in Sect. 4.}
The antiHiggs bundle ${\cal H}_N(\Si)$ is a cotangent bundle to the space
of connections $\nabla^{(1,0)}=\p+A$ in a vector bundle of rank $N$
over a Riemann curve
$\Si$. The cotangent vector (the antiHiggs field) is $\sln$ valued $(0,1)$-form
$\bar{\Phi}$. The symplectic form on ${\cal H}_N(\Si)$ is\\
$-\int_\Si\tr(d\bar{\Phi}\wedge dA)$.
The affinizations ${\cal R}^{(\ka)}/{\cal H}_N(\Si)$
are the space of connections \\$(\ka\bp+\bA,\p+A)$ with the
symplectic form $\int_\Si\tr(dA\wedge d\bA)$, where $\ka$ parameterizes the
affinizations. The elements of the space Conn$_{(\p+A)}({\cal L})$
giving rise to   ${\cal R}^{(\ka)}_{{\rm SL}(N)}/{\cal H}_N(\Si)$ are
\beq{na}
\nabla \Psi=\frac{\de \Psi}{\de A}+\ka \bA\Psi\,.
\eq

\subsection{Hamiltonian algebroids related to Lie algebroids}

Now we are ready to introduce an important subclass of Hamiltonian algebroids.
They are extensions of the Lie algebroids and share with them
SAJI (\ref {5.15}).

\begin{lem}
The anchor action (\ref{5.15i}) of the Lie algebroid ${\cal A}$ over $M$
can be lifted to the Hamiltonian action on ${\cal R}/T^*M$
 such that it defines the Hamiltonian algebroid ${\cal A}^H$ over ${\cal R}$.
The equivalence classes of these lifts
are isomorphic to $H^1({\cal A},M)$.
\end{lem}
{\sl Proof}.
Consider a small disk $U_\al\subset M$. The anchor (\ref{5.11}) has the form
\beq{5.15a}
\hat{\de}_{e^j}=\lan  b^j|\frac{\de}{\de  x}\ran   +c(x;e^j)\,.
\eq
Next, continue the action on ${\cal R}/T^*U_\al$.
We represent the affinization
as the space \\
Conn\,${\cal L}(M)=\{\nabla^p=\frac{\de}{\de  x}+p\,,~~ x\in U_\alpha\,,
~~p\in T^*M \}$.
Since ${\cal L}$ on $U_{\alpha}$ is trivialized we can identify the connections
with one-forms $p$.
Let $w\in TM$ and
$$
\nabla^p_w\Psi:=i_w\nabla^p\Psi=
\lan w|\frac{\de\Psi}{\de x}\ran+\lan  w|p\ran\Psi
$$
be the covariant derivative along $w$.
 To lift the action we use the
Leibniz
rule for the anchor action on the covariant derivatives:
$$
\hat\de_{e^j}(\nabla^p)_\al\Psi=\hat\de_{e^j}(\nabla^p_\al\Psi)-
\nabla^p_\al\hat\de_{e^j}\Psi-\nabla^p_{\hat\de_{e^j}\al}\Psi\,.
$$
It follows from (\ref{5.16i}) and (\ref{5.15a}) that
\beq{5.16a}
\hat{\de}_{e^j}p=-\frac{\de}{\de x}\lan p|b^j(x)\ran-\frac{\de}{\de x}c(x;e^j)\,.
\eq
Note that the second term is responsible for the pass from $T^*M$ to
the affinization ${\cal R}$, otherwise $p$ is transformed as a cotangent vector
(see (\ref{5.15i})).

The vector fields (\ref{5.15a}) are hamiltonian
 with respect
to the canonical symplectic form $\lan  dp|dx\ran   $ on ${\cal R}$.
The corresponding  Hamiltonians have the linear dependence on "momenta":
\beq{5.17}
h^j=\lan p|b^j(x)\ran   +c^j(x)\,.
\eq
Note that $h^j$ satisfies the Hamiltonian algebroid
property (\ref{5.12}), since $sc^j(x)=0$ (\ref{5.8}).

We have constructed the Hamiltonians locally and want to prove that this definition
is compatible with gluing
$U_\alpha$ and $U_\beta$. Note, that when we glue  ${\cal R}|_{U_\alpha}$
 and ${\cal R}|_{U_\beta}$
we shift fibers by $\frac{\de c_{\alpha\beta}}{\de x}$~:
$p_\alpha=p_{\beta}+ \frac{\de c_{\alpha\beta}}{\de x}$. Indeed,
 we glue the bundle ${\cal L}(M)$
restricted on $U_{\alpha\beta}$ by multiplication on  $\exp  c_{\alpha\beta}(x)$.
The connections are
transformed
by adding the logarithmic derivative of the transition functions.
On the other hand, $ \de_{\ve}c_{\alpha\beta}(x)=-c_\alpha(x;\ve)+c_\beta(x;\ve)$
(see (\ref{ch1})). So
$$
h_\alpha^j=\lan p_\alpha|b^j(x)\ran   +c_\alpha^j(x)=\lan p_{\beta}+
 \frac{\de c_{\alpha\beta}}{\de x}|b^j(x)\ran
 -\de_{\ve}c_{\alpha\beta}^j(x)+c_\beta^j(x;\ve)
 $$
$$
=\lan p_\beta|b^j(x)\ran   +c_\beta^j(x)=h_\beta^j\,,
$$
and the Hamiltonians become defined globally.

The exact cocycle $c_\alpha^j(x)=\de_{e^j}f_\alpha(x)$
shifts the momenta
$h^j=\lan p_\alpha+\frac{\de f_\alpha(x)}{\de x}|b^j(x)\ran$.

We rewrite the canonical transformations in the form
$$
\hat{\de}_{e^j}\Phi(p,x)=\de_{e^j}\Phi(p,x)+
\lan f^j|\frac{\de\Phi(p,x)}{\de p}\ran\,,~~
f^j=-\lan p|\frac{\de b^j(x)}{\de x}\ran   -\frac{\de c^j(x)}{\de x}\,.
$$
Thus, all nonequivalent
lifts of anchors from $M$ to ${\cal R}/T^*M$ are in one-to-one correspondence
with $H^1({\cal A},M)$.
Thereby, we have constructed the Hamiltonian
algebroid ${\cal A}^H$ over the principle homogeneous  space ${\cal R}$.
It has the same fibers and
the same structure functions $f^{jk}_i(x)$ as the underlying Lie algebroid ${\cal A}$
over $M$ and the bundle map $e^j\to h^j$ (\ref{5.17}).
$\Box$

\bigskip
Now investigate AJI (\ref{5.14}) in this particular case.
\begin{lem}
The Hamiltonian algebroids  ${\cal A}^H$ have the SAJI (\ref{5.15}).
\end{lem}
{\sl Proof}.

First note that the Lie algebroids we started with have the
SAJI (\ref{5.5}). The Hamiltonian algebroids
${\cal A}^H$ have the same structure functions
$f^{jk}_i(x)$ depending on coordinates on $M$ only. Consider
the general AJI (\ref{5.14}). It follows from (\ref{5.17})
that
$$
\{h^j,f^{nk}_i(x)\}=\lan b^j|\frac{\de f^{nk}_i(x)}{\de  x}\ran   =
\de_{e^j}f^{nk}_i(x)\,.
$$
The sum of the first two terms in (\ref{5.14}) coincides with
the SAJI (\ref{5.15}) in the underlying Lie algebroid ${\cal A}$, and therefore vanishes.
$\Box$

\subsection{Reduced phase space and its BRST description}

In what follows we shall consider Hamiltonian algebroids related to
Lie algebroids. Let $e^j$ be a basis of sections in $\G({\cal A}^H)$.
Then the Hamiltonians (\ref{5.17}) can be represented in the form
$h^j=\lan e^j|F(x)\ran   $, where $F(x)\in\G(({\cal A}^{H})^*)$ defines the {\em moment map}
$$
m: {\cal R}\rar\G(({\cal A}^{H})^*)\,,~~m(x)=F(x)\,.
$$
The coadjoint action ${\rm ad}^*_{\ve}$ in $\G(({\cal A}^{H})^*)$ is defined
in the usual way
$$
\lan \lb\ve, e^j\rb|F(x)\ran   =\lan e^j|{\rm ad}^*_{\ve}F(x)\ran \,  .
$$
One can fix a moment  $F(x)=m_0$ in $\G(({\cal A}^{H})^*)$.
 The reduced phase space is defined as the quotient
$$
{\cal R}^{red}=\{x\in {\cal R}|(F(x)=m_0)/G_0\}\,,
$$
where $G_0$ is  generated by the  transformations
${\rm ad}^*_\ve$ such that ${\rm ad}^*_\ve m_0=0$. In other words, ${\cal R}^{red}$
is the set of orbits of $G_0$ on the constraint surface $F(x)=m_0$.
The symplectic form $\om$ being restricted on $R^{red}$ is non-degenerate.

The BRST approach allows us to go around the reduction procedure by introducing
additional fields (the ghosts).
We shall construct the BRST complex for ${\cal A}^{H}$ in a similar way as the
Cartan-Eilenberg complex for the Lie algebroid ${\cal A}$. The BRST complex
is endowed with a Poisson structure and it
allows us to define the nilpotent BRST operator.

 Consider the dual bundle $({\cal A}^{H})^*$.
Its  sections $\eta \in\G(({\cal A}^{H})^*)$ are the anti-commuting (odd) fields called
 {\em the ghosts}.
Let $h_{e^j}=\lan \eta_j|F(x)\ran$, where $\{\eta_j\}$ is a basis in
$\G(({\cal A}^{H})^*)$ and $F(x)=0$ are the moment constraints, generating the
canonical algebroid action on ${\cal R}$.
 Introduce another type of odd variables ({\em the ghost momenta})
${\cal P}^j,~j=1,2,\ldots$ dual to the ghosts $\eta_k,~k=1,2,\ldots$
${\cal P}\in \G({\cal A}^{H})$.
 We attribute the ghost
 number one to the ghost fields gh$(\eta)=1$, minus one to the ghost momenta
gh$({\cal P})=-1$ and gh$(x)=0$ for $x\in {\cal R}$.
Introduce
the Poisson brackets in addition to the non-degenerate Poisson structure on ${\cal R}$
\beq{5.19}
\{\eta_j,{\cal P}^k\}=\de_j^k\,,~~\{\eta^j,x\}=\{{\cal P}_k,x\}=0\,.
\eq
Thus, all fields are incorporated in the graded Poisson superalgebra
$$
{\cal BFV}=\left(
\G(\wedge^\bullet  (({\cal A}^{H})^*\oplus{\cal A}^{H})
\right)\otimes C^\infty({\cal R})
=\G(\wedge^\bullet ({\cal A}^{H})^*)
\otimes\G(\wedge^\bullet {\cal A}^{H})\otimes C^\infty({\cal R})\,.
$$
({\em the Batalin-Fradkin-Volkovitsky (BFV) algebra}).

There exists a nilpotent operator $Q$ on the BFV algebra $Q^2=0,~gh(Q)=1$
({\em the BRST operator}) transforming ${\cal BFV}$ into the BRST complex.
The cohomology of the BRST complex give rise to the structure of
the classical reduced phase space ${\cal R}^{red}$. In some cases\\
$H^j(Q)=0,~j>0$ and $H^0(Q)=$ classical observables.

Represent the action of $Q$ as the Poisson brackets:
$$
Q\psi=\{\psi,\Om\},~~\psi,\Om\in {\cal BFV}\,.
$$
Due to the Jacobi identity for the Poisson brackets the nilpotency of $Q$
is equivalent to
$\{\Om,\Om\}=0$.
Since $\Om$ is odd, the brackets are symmetric.
 For generic Hamiltonian algebroid $\Om$
 can be represented as the expansion \cite{HT}
$$
\Om=h_\eta+\oh\lan \lb\eta,\eta'\rb|{\cal P}\ran   +...\,,~(h_\eta=\lan \eta|F\ran   )\,,
$$
where the higher order terms in ${\cal P}$ are omitted.
The highest order of ${\cal P}$
in $\Om$ is called {\em the rank} of the BRST operator $Q$.
If ${\cal A}$ is a Lie algebra defined along with its canonical
action on ${\cal R}$ then $Q$ has the rank one or less. In this case
the BRST operator $Q$ is the extension of the Cartan-Eilenberg operator
giving rise to the cohomology of ${\cal A}$ with coefficients in
$C^\infty({\cal R})$. Due to the Jacobi identity
the first two terms in the previous expression provide
the nilpotency of $Q$. It turns out that for the Hamiltonian algebroids ${\cal A}^H$
$\Om$ has the same structure as for  the Lie algebras
 though the Jacobi identity has additional terms.
\begin{theor}
The BRST operator $Q$ for the Hamiltonian algebroid ${\cal A}^{H}$
has the rank one:
\beq{5.20}
\Om=\lan \eta|F\ran   +\oh\lan \lb\eta,\eta'\rb|{\cal P}\ran \,.
\eq
\end{theor}
{\sl Proof}.\\
 Straightforward  calculations show that
$$
\{\Om,\Om\}=
\{h_{\eta_1},h_{\eta_2}\}+\oh\lan \lb\eta_2,\eta_2'\rb|F\ran
-\oh\lan \lb\eta_1,\eta_1'\rb|F\ran
$$
$$
+\oh\{h_{\eta_1},\lan \lb\eta_2\eta_2'\rb|{\cal P}_2\ran   \}
-\oh\{h_{\eta_2},\lan \lb\eta_1,\eta_1'\rb|{\cal P}_1\ran   \}+
\f1{4}\{\lan \lb\eta_1,\eta_1'\rb|{\cal P}_1\ran,\lan \lb\eta_2,\eta_2'\rb|{\cal P}_2\ran\}\,.
$$
The sum of the first three terms vanishes due to (\ref{5.12}).
The sum of the rest terms is the left hand side of the SAJI (\ref{5.15}).
The additional dangerous term may come from the Poisson brackets of
the structure functions
$\{\lb\eta_1,\eta_1'\rb,\lb\eta_2,\eta_2'\rb\}$. In fact, these
brackets vanish because the structure functions
do not depend
on the ghost momenta. Thus, the SAJI leads to the desired identity $\{\Om,\Om\}=0$.
$\Box$

\section{Hamiltonian algebroids and Poisson sigma-model}
\setcounter{equation}{0}

\subsection{Cotangent bundles to Poisson manifolds
as Lie algebroids}

Let $M$ be a Poisson manifold with Poisson bivector $\pi=\pi(\ve,\ve')$,
where    $\ve,\ve'$ are section of the bundle $T^*M$.
It is a skewsymmetric tensor, with the vanishing Schouten brackets (the Jacobi identity)
$[\pi,\pi]_S=0$. It means
in local coordinates $x=(x_1,\ldots,x_n)$
\beq{2.1}
\p_i\pi^{jk}(x)\pi^{im}(x)+{\rm c.p.}(j,k,m)=0\,.
\eq

The Poisson brackets are defined on the space of smooth functions
${\cal H}(M)$
$$
\{f(x),g(x)\}:=\lan df|\pi| dg\ran,~~df,dg\in T^*_xM\,.
$$
The Poisson bivector gives rise to  the  map
\beq{2.2a}
V^\pi:T^*M\to TM\,,~~
V^\pi_\ve=\lan\ve|\pi|\p\ran\,.
\eq
The corresponding Lie derivative ${\cal L}_{V_\ve}:=\de_\ve$
acts as
\beq{dx}
\de_\ve x=\pi(x)|\ve\ran\,.
\eq
In this way we obtain a map from the space of smooth functionals ${\cal H}(M)$
 to the space of the Hamiltonian vector fields $\G^H(TM)=\{V^\pi_\ve\}$
\beq{2.2}
f\to V_f= \lan df|\pi|\p\ran\,.
\eq
The Poisson brackets  can be rewritten as
$\{f(x),g(x)\}=-i_{V_{df}}dg$.

One can define the brackets on the one-forms $\ve,\ve'\in\G(T^*M)$
\beq{2.3}
\lb\ve,\ve'\rb=d\lan\ve|\pi(x)|\ve'\ran+\lan d\ve|\pi|\ve'\ran
+\lan\ve|\pi|d\ve'\ran\,.
\eq

\begin{lem}
$T^*M$ is a Lie algebroid ${\cal A}$ over the Poisson manifold $M$
with the Lie brackets (\ref{2.3}) and the anchor (\ref{2.2a}).
\end{lem}
{\sl Proof.}

It follows from the Jacobi identity (\ref{2.1}), that
the brackets (\ref{2.3}) is the Lie brackets and the
commutator of the vector fields  satisfies (\ref{5.1})
\beq{2.3a}
[ V_\ve,V_{\ve'}]=V_{\lb\ve,\ve'\rb}\,.
\eq
The property
(\ref{5.2}) follows from the definition of Lie brackets (\ref{2.3}).
$\Box$

\bigskip

The structure functions  of $T^*M$  is defined by the Poisson bivector
$$
f^{jk}_i(x)=\p_i\pi^{jk}(x)\,.
$$
This type of Lie algebroids was introduced in
Ref.~\cite{Ka},\cite{Fu}.

\begin{rem}
A linear space $M$ with the linear Poisson brackets $\pi^{jk}(x)=f^{jk}_ix^i$
can be identified with a Lie coalgebra $\gG^*$.
 Then the Hamiltonian vector field coming from the anchor $V^\pi$
is just the coadjoint action
\beq{co}
V^\pi_\ve\sim{\rm ad}^*_\ve \,.
\eq
\end{rem}

Consider a punctured
 disk $D^*=|z|\le 1$ and the space of the meromorphic maps\\
${\bf M} = \{X:D^*\to M,~X=X((z,z^{-1}))\}$.
Define the Lie algebroid ${\cal A}_{\bf M}$ over the space
${\bf M}$ with the brackets (\ref{2.3}) on  the sections of $X^*(T^*M)$.
For simplicity we do not change the notion of the anchor action
\beq{7.4b}
\de_\ve X=\pi(X)|\ve\ran\,.
\eq

\subsection{Poisson sigma-model and Hamiltonian algebroids}

The Poisson sigma-model is a way to construct a Hamiltonian
algebroid related to the Lie algebroid  ${\cal A}_\bfM$.
The manifold $M$ serves as the target space for the Poisson sigma model.
The space-time  is the disk $D^*$.

Consider  the one-form $\xi$ on $D^*$ taking values in the pull-back by
$X$ of the cotangent bundle $T^*M$, or the affinization over $T^*M$.
Endow the space of fields $(X,\xi)$ with
the canonical symplectic form
\beq{7.2a}
\om=\f1{2\pi}\oint\lan D X \we D\xi \ran.
\eq

The canonical transformations of $\om$  is
represented by (\ref{7.4b}) and according with (\ref{5.16a}) by
\beq{7.5}
\hat{\de}_\ve\xi=\frac{\de}{\de X}c(X,\ve)
+\lan \ve|\frac{\de}{\de X}\pi|\xi\ran=
-\bp\ep+\lan\ve|\frac{\de}{\de X}\pi|\xi\ran\,,
\eq
where  $c(X,\ve)$ is the anti-holomorphic one-cocycle  from $H^1({\cal A},\bfM)$
\beq{7.7}
c(X,\ve)=-\f1{2\pi}\oint\lan\ve|\bp X\ran\,.
\eq
These transformations are generated by the first class constraints
 \beq{7.3}
F:=\bp X+\pi(X)|\xi\ran=0\,.
\eq

The action (\ref{7.5}) amounts to the lift of the anchor action from $\bfM$
to
the affinization ${\cal R}/T^*\bfM$ over $T^*\bfM$
 by means of the cocycle (\ref{7.7}) in accordance with Lemma 2.1.
The canonical transformations of a smooth functionals on ${\cal R}$
 are the Hamiltonian transformation
\beq{anc}
\hat{\de}_{\ve} f(X,\xi)=\{h_{\ve},f(X,\xi)\}\,.
\eq
Here the Poisson brackets are inverse to the symplectic form $\om$
(\ref{7.2a})
and
\beq{7.8}
h_{\ve}=\oint\lan\ve| F\ran\,.
\eq
(see (\ref{5.17})).
Again, due to the cocycle property,
$\{h_\ve,h_{\ve '}\}=h_{\lb\ve,\ve '\rb}$.

Summarizing, we have defined the symplectic manifold
 ${\cal R}\{\xi,X\}$
and the Hamiltonian algebroid
${\cal A}^H_{\cal R}$ over ${\cal R}$ with the sections
$\ve\in\G({\cal A}^H_{\cal R})$ and the anchor (\ref{anc}),(\ref{7.8}).

Following our approach we interpret the constraints  (\ref{7.3}) as the consistency
conditions for a linear system.
Let $\psi_1,\psi_2$ be sections of $X^*(T^*M)$ and $X^*(TM)$
correspondingly, and $B$ is a family of the linear maps
$X^*(T^*M)$ to $X^*(TM)$
\beq{7.4c}
B(X)=\la+\pi (X)\,,~~B(X)\psi_1=\psi_2\,.
\eq
Consider the linear system
\beq{7.4a}
B(X)\psi_1=0\,.
\eq
The second equation is
\beq{7.5a}
A^*\psi_1=0\,,
\eq
where $A$ is the linear map
$A:X^*(T^*M)\to X^*(T^*M)$
\beq{7.5c}
A=-\bp+|\frac{\de}{\de X}\pi|\xi\ran\,,
\eq
and $A^*:X^*(TM)\to X^*(TM)$
$$
A^*=-\bp-|\frac{\de}{\de X}\pi|\xi\ran\,.
$$
\begin{lem}
Let the Poisson bivector satisfies the non-degeneracy condition:\\
 the matrix $a_i^j=(\frac{\de}{\de X^i}\pi ^{jm}$ is non-degenerate
on $M$ for some $m$.\\
 Then the constraints  (\ref{7.3}) are the consistency
conditions for (\ref{7.4a}) and (\ref{7.5a}).
\end{lem}
{\sl Proof.}
The consistency condition of these equations is the operator equation
$$
BA-A^*B=0.
$$
After substitution in it the expressions for $A,A^*,B$ and applying the Jacobi identity
(\ref{2.1}) one comes to the equality
$$
(\bp X^j+\pi^{is}\xi_s)\p_i\pi^{jm}(\psi_1)_m=0.
$$
The later is equivalent to the constraint equation (\ref{7.3}) if $\pi$ is
non-degenerate in the above sense. $\Box$

\subsection{BRST construction}

Consider a smooth functional  $\Psi(X)$ on $\bfM$
annihilating by the anchor action
\beq{LB}
\hat{\de}_{\ve}\Psi(X):=\f1{2\pi}\oint\lan\ve|
\pi(X)d\Psi(X)\ran+
\left(\f1{2\pi}\oint\lan\ve|\bp X\ran
\right)\Psi(X)=0.
\eq
Let $G$ be the Lie groupoid corresponding to the Lie algebroid ${\cal A}_\bfM$.
Consider the space of orbits $\ti{\bfM}=\bfM/G$
and the line bundle ${\cal L}(\ti{\bfM})$ over
$\ti{\bfM}$ with
 the space of sections $\G({\cal L})=\{\Psi(X)\}$ (\ref{LB}).
The Hilbert space $L^2(\G({\cal L}))$ arises in the prequantization
of the symplectic quotient\\
${\cal R}^{red}={\cal R}//G^H=\{F=0\}/G^H$,
where $G^H$ is the Hamiltonian groupoid.

The quantization of ${\cal R}^{red}$ can be performed by the BRST technique.
The classical BRST complex is the set of fields
$$
\bigwedge{}^\bullet\left
(\G(X^*(TM))
\oplus\G(X^*(T^*M))
\right)\otimes C^\infty({\cal R})\,,
$$
where the first component is the space of sections of the anticommuting
variables $\eta$ dual to the gauge generators $\ve$, the second component
is the space of their momenta ${\cal P}$.
Theorem 2.1 states that the BRST operator has the rank one
$$
\Om=\f1{2\pi}\oint\lan\eta |F\ran+
\f1{\pi}\oint\lan\lb\eta,\eta\rb|{\cal P}\ran\,.
$$
It means that the nonlinear deformation of the Poisson bivector on $M$
does not affect the Lie algebraic form of $\Om$.
This form of $\Om$ for the Poisson sigma model
was established in \cite {SS}.

\section{Two examples of Hamiltonian algebroids with
Lie algebra symmetries}
\setcounter{equation}{0}

In this section we consider two examples, where the spaces of sections
of the Hamiltonian
algebroids are just  Lie algebras of hamiltonian
vector fields and therefore the symmetries are the standard Lie
symmetries.
Nevertheless, they are in much the same
as in the  algebroid cases.
In both examples we cast the known constructions in the form suitable
to our approach.

Let $\Si_{g,n}$ be a Riemann curve of
 genus $g$
with $n$ marked points. The first examples is the moduli
space of flat bundles over $\Si_{g,n}$. It will become clear later,
that it is
an universal system containing hidden algebroid symmetries.
 The second example is
the projective structures (${\cal W}_2$-structures) on $\Si_{g,n}$.
Their generalization is the ${\cal W}_N$-structures, where
the symmetries are
defined by a nontrivial Hamiltonian  algebroid, will be considered
in next Sections.

\subsection{Flat bundles  with the regular singularities}

Consider a ${\rm SL}(N,{\bf C})$ holomorphic bundle  $E$ over
 $\Si_{g,n}$.
Let $D^*$ be a small punctured disk embedded to $\Si_{g,n}$
with a local coordinate $z$.
Locally  the derivatives
 $d': E\to E\otimes\Om^{(1,0)}(\Si_{g,n})$,
$d'': E\to E\otimes\Om^{(0,1)}(\Si_{g,n})$ take the form
\beq{8.0}
d'=\p +A\,, ~~d''=\bp\,.
\eq

Let $M_{SL_N}(D^*)=\{d'=k\p+A\}$ be the set of
derivatives restricted to $D^*$.  This set has the structure
of the affine Lie coalgebra
$\hat{L}^*(\sln)$ with the Lie-Poisson brackets on the space of smooth functionals
$$
\{f(A),g(A)\}=
\oint\tr([df(A),dg(A)]A)+
\oint\tr(df\p (dg) )\,,
$$
where $df(A)\in L(\sln)$ is the variation of $f(A)$.
Thereby $M_{SL_N}(D^*)$ can be considered as
 the base of the Lie algebroid ${\cal A}_{SL_N}(D^*)$
 (see Remark 3.1).
The space of sections of the algebroid is the Lie algebra
${\cal G}_{SL_N}(D^*)=L(\sln)$
of  the gauge transformations
\beq{8.1}
\de_\ve A=\p\ve+[A,\ve]\,.
\eq

Though the Poisson structure
is defined only on $D^*$ the algebroid
can be defined over $\Si_{g,n}$, since the gauge algebra
and the anchor action (\ref{8.1}) are well defined globally.
We denote by ${\cal G}_{SL_N}$  the algebra
of the smooth global gauge transformations.
To come to the global description
we assume that $A$ has first order holomorphic poles at the marked points
\beq{8.0a}
A|_{z\to x_a}=\frac{A_a}{z-x_a}\,.
\eq
In addition, we consider a collection $P$ of $n$  elements from  the
Lie coalgebra
$$
P=\left\{\bfp=(p_1,\ldots,p_a,\ldots,p_n), ~~p_a\in{\rm sl}^*(N,{\mC})\right\}\,.
$$
endowed with the Lie-Poisson structure
$
\{f(p_a),g(p_b)\}=\de_{ab}\tr([ df,dg]p_a)$.
We assume that the gauge transformations at the marked points are nontrivial
\beq{8.0b}
\ve|_{z\to x_a}=r_a+O(z-x_a)\,, ~~r_a\neq 0\,.
\eq
The gauge algebra ${\cal G}_{SL_N}$ acts on $P$ by the evaluation maps
\beq{8.1a}
\de_\ve p_a=[p_a,r_a]\,,~~
\ve\in {\cal G}_{SL_N}\,.
\eq
In this way we define a trivial Lie algebroid
${\cal A}_{SL_N} ={\cal G}_{SL_N}\times M_{SL_N}$
over
$M_{SL_N}=\{d',P\}$  with the anchor map (\ref{8.1}), (\ref{8.1a}).

The cohomology
$H^i({\cal A}_{SL_N})=H^i({\cal G}_{SL_N},M)$
are the standard cohomology
of the gauge algebra ${\cal G}_{SL_N}$ with the cochains taking values
in functionals on $M$.  There is a nontrivial one-cocycle
\beq{8.2}
c(A,\bfp;\ve)=\int_{\Si_{g,n}}\tr\ve
\left(
\bp A-2\pi i\sum_{a=1}^n\de(x_a)p_a
\right)=
\lan\ve|\bp A\ran-2\pi i\sum_{a=1}^n\tr(r_a\cdot p_a)
\eq
representing an element of $H^1({\cal A}_{SL_N})$.
 This cocycle provides a nontrivial
extension of the anchor action (see (\ref{5.11}))
\beq{8.2a}
\hat{\de}_\ve f(A,\bfp)
=\lan\ve
|\bp A-\p (df(A))+[df(A),A]\ran
-2\pi i\sum_{a=1}^n\tr(r_a p_a) \,.
\eq

Next consider $2g$ contours $\ga_\al,~(\al=1,\ldots,2g)$
 generating $\pi_1(\Si_{g})$. The contours determine the 2-cocycles
\beq{8.3}
c_\al(\ve_1,\ve_2)=\int_{\ga_\al}\tr(\ve_1\p\ve_2)\,.
\eq
The cocycles (\ref{8.3}) lead to $2g$ central extensions
$\hat{\cal G}_{SL_N}$ of ${\cal G}_{SL_N}$
$$
\hat{\cal G}_{SL_N}={\cal G}_{SL_N}\oplus_{\al=1}^{2g}{\bf C}\La_\al\,,
$$
$$
[(\ve_1,\sum_{\al}k_{1,\al}),(\ve_2,\sum_{\al}k_{2,\al})]_{c.e.}=
\left([\ve_1,\ve_2],\sum_{\al}c_\al(\ve_1,\ve_2)\right)\,.
$$

To define the corresponding Poisson sigma-model we
consider the cotangent bundle $T^*E$.
The conjugate to
$\p +A$ variables are the one-forms
$\bar{\Phi}\in\Om^{(0,1)}(\Si_{g,n},{\rm sl}(N,{\mC}))$
 -- the antiHiggs field.
In fact, we shall consider the affinization
${\cal R}^0_{SL_N}=Aff(T^*E)$ over
$T^*E$ provided
by the cocycle (\ref{8.2}). We have already mentioned that
 the role of momenta plays by  the holomorphic connection $\bp+\bA$ (\ref{na}).
 We put $\ka=1$. The symplectic form on ${\cal R}^0_{SL_N}$ is
$$
\om^0= \lan DA\wedge D\bA\ran\,.
$$

Consider the contributions of the marked points. We define there the symplectic manifold
$$
(T^*G_1,\ldots,T^*G_n)\,.
$$
The dual variables to $p_a$ at the marked points are $n$ elements
 $g_a\in\SLN$
and the symplectic form  is
\beq{8.4c}
\sum_{a=1}^n\om_a=\sum_{a=1}^nD\tr(p_a\wedge g_a^{-1}Dg_a)\,.
\eq
Here
$\om_a$
is the canonical symplectic form on $T^*G_a\sim T^*\SLN$.
We pass from $T^*G_a$ to the coadjoint orbits
$$
{\cal O}_a=\{p_a=g_a^{-1}p_a^{(0)}g_a~|~p_a^{(0)}=\di(\la_{a,1},\ldots,\la_{a,N}),~
\la_{a,j}\neq\la_{a,k},~g_a\in{\rm SL}(N,{\bf C})\}\,,
$$
The orbits are the symplectic quotient
${\cal O}_a\sim \SLN\backslash\backslash  T^*G_a$ with respect to the action\\
$g_a\to f_ag_a,~f_a\in\SLN$.
The form $\om_a$ coincides on ${\cal O}_a$ with the Kirillov-Kostant form
$\om_a=D\tr(p^{(0)}_a Dg_ag_a^{-1})$. The orbits ${\cal O}_a$ are
affinizations $Aff(T^*Fl_a(N))$ over
the cotangent bundles $T^*Fl_a(N)$ to the flag varieties $Fl_a(N)$.

Eventually we come
 to the symplectic manifold
$$
{\cal R}_{SL_N}=({\cal R}^0_{SL_N};{\cal O}_1,\ldots,{\cal O}_n)\sim Aff(T^*E);Aff(T^*Fl_1),\ldots,Aff(T^*Fl_n)\,,
$$
\beq{8.4a}
\om=\om^0+\sum_{a=1}^n\om_a=\lan DA\wedge D\bA\ran+\sum_{a=1}^nD\tr(p^0_a\wedge Dg_a g_a^{-1})\,.
\eq

 According to
(\ref{5.17}) the lift of the anchor (\ref{8.1}) to ${\cal R}_{N}$,
defined by the cocycle $c(A;\ve)$ (\ref{8.2}) leads  to the Hamiltonian
$$
h_\ve=
\lan\ve|F(A,\bA)-2\pi i\sum_{a=1}^n\de(x_a)p_a)\ran,~~
F(A,\bA)=\bp A-\p\bA+[\bA,A]\,.
$$
The Hamiltonian generates the canonical vector fields (\ref{8.1}) (\ref{8.1a}) and
$$
\hat{\de}_\ve \bA=\bp\ve+[\bA,\ve],~~\hat{\de}_\ve g_a=g_ar_a,
$$
(see (\ref{5.16a})).
The global version of this transformations is the gauge group
$G_{SL_N}$ acting on ${\cal R}_{SL_N}$.
The flatness condition
\beq{8.5a}
m:=F(A,\bA)-2\pi i\sum_{a=1}^n\de(x_a)p_a=0
\eq
is the moment constraint with respect to this action. This equation means
that the residues $A_a$ of $A$ in the marked points (\ref{8.0a}) coincide with
$p_a$.
The flatness is the compatibility condition for the linear
system
\beq{8.5c}
\left\{
\begin{array}{c}
(\p +A)\psi=0\,,\\
(\bp +\bA)\psi=0\,,
\end{array}
\right.
\eq
where $\psi\in\Om^0(\Si_{g,n},{\rm Aut} E)$. The second equation
describes the deformation of the holomorphic structure of the bundle $E$
 (\ref{8.0}).

The moduli space ${\cal M}^{flat}_N$ of flat $\SLN$-bundles
is the symplectic quotient
${\cal R}_{SL_N}//G_{SL_N}$. It has  dimension
\beq{8.5b}
\dim {\cal M}^{flat}_N=2(N^2-1)(g-1)+N(N-1)n\,,
\eq
where the last term is the contribution of the coadjoint orbits ${\cal O}_a$.
Let
$\ti{M}_{SL_N}=M_{SL_N}/G_{SL_N}$ be the set of the
gauge orbits acting on the base space $M_{SL_N}=\{d',P\}$.
Consider smooth functionals $\Psi(A,\bfp)$ annihilated by the
anchor action $\hat{\delta}_\varepsilon\Psi(A,\bfp)=0$.
These functionals generate the space of sections of the linear bundle
${\cal L}(\ti{M}_{SL_N})$ we
discussed before (\ref{5.11b}). It  is the determinant bundle
 $\det(\p+A)$ \cite{ADW,H1}.
The prequantization of ${\cal M}^{flat}_N$ is defined in the Hilbert space
 $L^2(\G({\cal L}(\ti{M}_{SL_N}))$.

On the other hand ${\cal M}^{flat}_N$ can be described by the
cohomology $H^k(Q)$ of the BRST operator $Q$ which we are going to define.
Let $\eta$ be the  dual to $\ve$ fields (the ghosts)
and ${\cal P}$ are their momenta ${\cal P}\in\Om^{(1,1)}(\Si_{g,n},{\rm End}E)$.
 Consider the algebra
$$
C^\infty({\cal R}_{N})\otimes\wedge^\bullet
 \left({\cal G}_{ SL_N})\oplus {\cal G}^*_{ SL_N}\right)\,.
$$
Then the BRST operator $Q$ acts on functionals on this algebra as
$$
Q\Psi(A,\bA,\eta,{\cal P})=\{\Om,\Psi(A,\bA,\eta,{\cal P})\}\,,
$$
 where
$$
\Om=\lan \eta|F(A,\bA)\ran   +\oh\lan [\eta,\eta']|{\cal P}\ran   \,,
$$
where res$A|_{x_a}=p_a$.

\subsection{Projective structures on $\Si_{g,n}$}

Consider the space $M_2$ of
projective connections  on $\Si_{g,n}$. Locally  on a punctured disk
$D^*$ the space $M_2$ is  represented by
the set $M_2(D^*)$ of the second order differential operators $\p^2-T$.
The later is the Poisson manifold
with the linear brackets
\beq{vir}
\{T(z),T(w)\}=(\frac{1}{2}\p^3+2T\p+\p T)\de(z-w)\,,
\eq
 where $ \de(z-w)=\sum_{k\in\mZ}z^kw^{-k-1}$ is the delta function
 on the contour $|z|=1$.

 The dual space to $M_2(D^*)$  is the  Virasoro algebra $Vir={\cal G}_1(D^*)=\{\ve,\bfc\}$
  $(\ve=\ve(z,\bz)\frac{\p}{\p z})$. The commutation relations can be read off
 from the Poisson brackets  (see (\ref{2.3}))
\beq{8.7b}
[\ve_1,\ve_2]=(\ve_1\p\ve_2-\ve_2\p\ve_1, \frac{1}{2}\oint\ve_1\p^3\ve_2)\,.
\eq

 The coadjoint action of $Vir$ on $M_2(D^*)$
\beq{8.7}
\de_{\ve}T(z,\bz)=-\ve\p T-2T\p\ve-\frac{1}{2}\p^3\ve\,.
\eq
defines the anchor in the trivial algebroid ${\cal A}_{2}(D^*)={\cal G}_1(D^*)\oplus M_2(D^*)$.

The algebroid (\ref{8.7b}), (\ref{8.7}) can be  defined
globally over the space $M_2$.  The section of the algebroid are the chiral
vector fields $\ve\sim \ve(z,\bz)\frac{\p}{\p z}$ on $\Si_{g,n}$.  Its central
extension is defined by the contour in $D^*$ (\ref{8.7b}).
 We call this algebra ${\cal G}_1$.
We include in the definition the contribution of the marked points.
Assume that the projective connections $T$ have
poles at the marked points $x_a,(a=1,\ldots,n)$
  up to the second order:
\beq{8.8a}
T|_{z\rar x_a}\sim\frac{T^a_{-2}}{(z-x_a)^2}+\frac{T^a_{-1}}{(z-x_a)}+\ldots\,,
\eq
and the vector fields have the first order holomorphic nulls at the marked points
\beq{8.8b}
\ve|_{z\rar x_a}=r_a(z-x_a)+o(z-x_a),~~r_a\neq 0\,.
\eq
We denote this trivial algebroid bundle  ${\cal A}_2$.

Consider the cohomology $H^\bullet ({\cal A}_2)\sim H^\bullet ({\cal G}_1,M_2)$.
Due to (\ref{8.7}) and (\ref{8.8b})  $\de_{\ve}T^a_{-2}=0$ and thereby
 $T^a_{-2}$ in (\ref{8.8a}) represents an element
from $H^0({\cal A}_2)$.

The anchor action (\ref{8.7}) can be extended by the one-cocycle $c(T;\ve)$
representing a nontrivial element $c(\ve;T)$ of $H^1({\cal A}_2)$
\beq{8.9a}
c(T;\ve)=\int_{\Si_{g,n}}\ve\bp T=\lan\ve|\bp T\ran\,,
\eq
\beq{8.9}
\hat{\de}_{\ve}f(T)=\lan\de_\ve T|df(T)\ran+c(T;\ve)\,.
\eq
The contribution of the marked point in (\ref{8.9a}) is $2\pi ir_aT_{-2}^a$.

There exist  $2g$ nontrivial two-cocycles defined by the integrals
over non contractible contours $\ga_\al$:
$$
c_\al (\ve_1,\ve_2)=\oint_{\ga_\al}\ve_1\p^3\ve_2.
$$
The cocycles give rise to the central extension $\hat{\cal G}_1$ of the Lie algebra
of the first order differential operators on $\Si_g$.

The affinization ${\cal R}_2$ over the cotangent bundle $T^*M_2$ has
the Darboux coordinates
$T$ and $\mu$,
where $\mu\in\Om^{(-1,1)}(\Si_{g,n})$ is the Beltrami differential.
The anchor (\ref{8.7}) is lifted to ${\cal R}_2$ as
\beq{8.11}
\de_\ve\mu=-\ve\p\mu +\mu\p\ve+\bp\ve\,,
\eq
where the last term occurs due to the cocycle (\ref{8.9a}).
We specify the dependence of $\mu$ on the positions of
the marked points in the following  way. Let ${\cal U}'_a$ be
 neighborhoods
 of the marked points $x_a,~(a=1,\ldots,n)$
such that ${\cal U}'_a\cap{\cal U}'_b=\emptyset$ for $a\neq b$.
Define a smooth function $\chi_a(z,\bz)$
\beq{cf}
\chi_a(z,\bz)=\left\{
\begin{array}{cl}
1,&\mbox{$z\in{\cal U}_a$ },~{\cal U}'_a\supset{\cal U}_a\\
0,&\mbox{$z\in\Si_g\setminus {\cal U}'_a$}\,.
\end{array}
\right.
\eq
Due to (\ref{8.11}) at the neighborhoods of the marked points $\mu$ is defined
 up to the term\\ $\bp(z-x_a)\chi(z,\bz)$.
Then $\mu$ can be represented as
\beq{mu}
 \mu=\sum_{a=1}^n[t^{(1)}_{0,a}+t^{(1)}_{1,a}(z-x_a)+\ldots]\mu^0_a\,,~~
\mu^0_a=\bp \chi_a(z,\bz)\,,~~(t_{0,a}=x_a-x_a^0)\,,
\eq
where only $t_{0,a}$ can not be removed by the gauge transformations (\ref{8.8b}), (\ref{8.11}).
The symplectic form on ${\cal R}_2$ is
$$
\om=\int_{\Si_{g,n}}dT\wedge d\mu=\lan dT\wedge d\mu\ran\,.
$$
For rational curves $\Si_{0,n}$ it takes the form
\beq{8.14}
\om=dT^a_{-2}\wedge dt_{1,a}+dT^a_{-1}\wedge dt_{0,a}\,.
\eq
\begin{rem}
The space ${\cal R}_2$ is the classical phase space
of the $2+1$-gravity on $\Si_{g,n}\times I$ \cite{Ca}. In fact,
$\mu$ is related to the conformal class of metrics on $\Si_{g,n}$
and plays the role of a coordinate, while $T$ is a momentum.
In our construction $\mu$ and $T$ interchange their roles.
\end{rem}

The Hamiltonian of the canonical transformations (\ref{8.7}), (\ref{8.11})
has the form
\beq{8.12}
h_\ve=\lan\ve| F(T,\mu)\ran\,,
\eq
$$
F(T,\mu)=
(\bp+\mu\p+2\p\mu)T-\frac{1}{2}\p^3\mu\,.
$$

The Hamiltonian defines the moment map  $m:{\cal R}_2\rar {\cal G}^*_1$
\beq{8.13}
m=(\bp+\mu\p+2\p\mu)T-\frac{1}{2}\p^3\mu\,,
\eq
where ${\cal G}^*_{1}$ is the dual to ${\cal G}_{1}$ space
of distributions  of $(2,1)$-forms on $\Si_{g,n}$.
As it follows from
(\ref{8.8b}) in the neighborhoods of the marked points the elements
 $y\in{\cal G}^*_{1}$ takes the form
\beq{8.13a}
y\sim b_{1,a}\p\de(x_a)+b_{2,a}\p^2\de(x_a)+\ldots\,.
\eq
We put $m$ equal
\beq{8.15}
m=-\sum_{a=1}^nT^a_{-2}\p\de(x_a)\,,~~(m=F(T,\mu))\,.
\eq
The algebra  ${\cal G}_1$  preserves $m:{\rm ad}_\ve^*m=m$
for any $\ve$.
Thus, in contrast with the previous example, we have trivial coadjoint orbits at
the marked points. Since $T^a_{-2}$ are fixed, the dynamical parameters  are
$(t_{0,a},T^a_{-1})$ that contribute in the symplectic structure (\ref{8.14}).
Let $\psi$ be a $(-\oh,0)$ differential. Then (\ref{8.15})
is the compatibility condition for the linear system
\beq{8.16}
\left\{
\begin{array}{l}
(\p^2-T)\psi=0\,,\\
(\bp+\mu\p -\oh\p\mu) \psi=0\,.
\end{array}
\right.
\eq
It follows from the second equation that the Beltrami differential $\mu$ provides
 the deformation of complex structure on $\Si_{g,n}$. Note, that we started
from the first equation defining the projective
connection  and $\bp\psi=0$ on $M_2$. The last equation is modified
after the passage from $M_2$ to  ${\cal R}_2$.

The moduli space ${\cal W}_2$ of projective structure on $\Si_{g,n}$ is
the symplectic quotient of ${\cal R}_2$ with respect to the action of $G_1$,
where $G_1$ is the group corresponding to the algebra ${\cal G}_1$
$$
{\cal W}_2={\cal R}_2//G_1=\{F(T,\mu)-m=0\}/G_1\,.
$$
It has dimension $6(g-1)+2n$. To quantize ${\cal W}_2$
 one can consider the quotient space $\ti{M}_2=M_2/G_1$.
 The space of sections of
the linear bundle ${\cal L}(\ti{M}_2)$ is defined as the space of
functionals $\{\Psi(T)\}$ on $M_2$ that satisfy the invariance condition\\
$\hat{\delta}_\ve\Psi(T)=0$.
The linear bundle ${\cal L}(\ti{M}_2)$ is the determinant line bundle
$\det(\p^2-T)$ considered in \cite{Za,Mat}.

The tangent space ${\cal T}_2$ to ${\cal W}_2$ is isomorphic
to the cohomology $H^0$ of the BRST complex. It is generated by
the fields $T,\mu\in {\cal R}_2$, the ghosts fields $\eta$ dual to the
vector fields $\ve$ acting via the anchor (\ref{8.7}),(\ref{8.11})
on ${\cal R}_2$ and the ghosts momenta ${\cal P}$.
The BRST operator $Q$ is defined by $\Om$
$$
\Om=\int_{\Si_{g,n}}\eta F(T,\mu)+\oh\int_{\Si_{g,n}}[\eta,\eta']{\cal P}.
$$
The first term is just the Hamiltonian (\ref{8.12}), where the vector fields are
replaced by the ghosts.

\section{Hamiltonian algebroid structure in ${\cal W}_3$-gravity}
\setcounter{equation}{0}

Now consider the concrete example of the general construction with nontrivial
algebroid structure.
It is the ${\cal W}_N$-structures on $\Si_{g,n}$ \cite{P,BFK,GLM}
which generalize the projective structures described in previous Section.
In this Section we consider in details the ${\cal W}_3$-structures.

\subsection{$\SLN$-opers}

Opers are $G$-bundles over Riemann curves with additional structures
\cite{Tel,BD}.
Let $E_N$ be a  $\SLN$-bundle over $\Si_{g,n}$. It is a $\SLN$-{\em oper} if
 there exists a flag filtration
$E_N\supset  \ldots\supset   E_1\supset    E_0=0$ and a covariant derivative,
 that acts as
$\nabla:~E_j\subset E_{j+1}\otimes\Om^{(1,0)}(\Si_{g,n})$.
Moreover, $\nabla$ induces an isomorphism
$E_j/E_{j-1}\to E_{j+1}/E_{j}\otimes\Om^{(1,0)}(\Si_{g,n})$.
It means that locally
\beq{opN}
\nabla=\p-
\left(
\begin{array}{cccccc}
0   & 1 & 0 & \ldots&   & 0 \\
0   & 0 & 1 & \ldots&   &   \\
    &   &   & \cdot &   &   \\
 0  & 0 &   &       & 0&  1  \\
W_N& W_{N-1}&\ldots&&W_2&0
\end{array}
\right)\,.
\eq
In other word we define the $N$-order differential operator on $\Si_{g,n}$
\beq{dopN}
L_N=\p^N-W_2\p^{N-2}\ldots-W_N~:
~\Om^{(-\frac{N-1}{2},0)}(\Si_{g,n})\to\Om^{(\frac{N+1}{2},0)}(\Si_{g,n})
\eq
with vanishing subprinciple symbol. The $\GL$-opers come from the $\GL$-bundles
and have the additional term $-W_1\p^{N-1}$ in (\ref{dopN}).
We assume that in  neighborhoods  of the marked points
the coefficients $W_j$ behave as
\beq{6.4a}
W_j|_{z\rar x_a}\sim
W^a_{-j}(j){(z-x_a)^j}+W^a_{-j}(j-1){(z-x_a)^{j-1}}+
\ldots\,.
\eq

In this section we consider $\SLt$-opers and postpone the general case to next Section.
It is possible to choose $E_1=\Om^{-1,0}(\Si_{g,n})$.
Instead of (\ref{opN}) we have
\beq{6.1}
\nabla=\p -\thmat{0}{1}{0}{0}{0}{1}{W}{T}{0}\,,
\eq
and the third order differential operator
\beq{6.2}
L_3=\p^3-T\p-W:~\Om^{(-1,0)}(\Si_{g,n})\to\Om^{(2,0)}(\Si_{g,n})\,.
\eq
According with (\ref{6.4a})
\beq{6.3}
T|_{z\rar x_a}\sim\frac{T^a_{-2}}{(z-x_a)^2}+\frac{T^a_{-1}}{(z-x_a)}+\ldots\,.
\eq
\beq{6.4}
W|_{z\rar x_a}\sim
\frac{W^a_{-3}}{(z-x_a)^3}+\frac{W^a_{-2}}{(z-x_a)^2}+\frac{W^a_{-1}}{(z-x_a)}
+\ldots\,.
\eq

\subsection{Local Lie algebroid over SL$(3,\mC)$-opers}

Consider the set $M_3(D^*)=\{L_3\}$ of $\SLt$-opers on a punctured disk $D^*$.
This set is a Poisson manifold with respect to the AGD brackets
\beq{6.9a}
\{T(z),T(w)\}
=\left(-2\ka\p^3+2T(z)\p+\p T(z)\right)\de(z-w)\,,
\eq
\beq{6.10a}
\{T(z),W(w)\}=\left(\p^4-T(z)\p^2+3W(z)\p-\p W(z)\right)\de(z-w)\,,
\eq
\beq{6.12a}
\{W(z),W(w)\}=
\eq
$$
+\left(\frac{2}{3}\p^5-\frac{4}{3}T(z)\p^3-
2\p T(z)\p^2+\left(\frac{2}{3}T(z)^2-2\p^2T(z)+2\p W(z)\right)\p
\right.
$$
$$
\left. +\left(\p^2W(z)-\frac{2}{3}\p^3T(z)+\frac{2}{3}T(z)\p T(z)\right)
\right)\de(z-w)\,.
$$

It means that $M_3(D^*)$ is a base of the local Lie algebroid
${\cal A}_3(D^*)\sim T^*M_3(D^*)$.
To define the space of its sections we consider
the dual space  $M^*_3 (D^*)$ of the space of second order differential operators on $D^*$ with a central extensions
$$
M^*_3 (D^*)=
\{ (\ve^{(1)}\frac{d}{dz}+ \ve^{(2)}\frac{d^2}{dz^2}),c\}\,.
$$
It comes from the pairing
\beq{pa}
\oint_{|z|=1}(\ve^{(1)}T+ \ve^{(2)}W)
\eq
and the central element $c$ is dual to the highest order coefficient
in(\ref{6.3}) which we put equal to $1$.
The Lie brackets on $ T^*M_3 (D^*)$ are determined by means of
the AGD Poisson structure following  (\ref{2.3})
\beq{6.6}
\lb\ve^{(1)}_1,\ve^{(1)}_2\rb=
\left((\ve^{(1)}_1\p\ve^{(1)}_2-\ve^{(1)}_2\p\ve^{(1)}_1) \frac{d}{dz}\,,\,
-2\oint\ve^{(1)}_1\p^3\ve^{(1)}_2\right)\,.
\eq
\beq{6.7}
\lb\ve^{(1)},\ve^{(2)}\rb=\left((
-\ve^{(2)}\p^2\ve^{(1)}) \frac{d}{dz}+( -2\ve^{(2)}\p\ve^{(1)}+\ve^{(1)}\p\ve^{(2)}) \frac{d^2}{dz^2})\,,\,
\oint\ve^{(1)}_1\p^4\ve^{(2)}_2\right)\,.
\eq
\beq{6.8}
\lb\ve^{(2)}_1,\ve^{(2)}_2\rb=\left(
\frac{2}{3}[\p(\p^2-T)\ve^{(2)}_1]\ve^{(2)}_2 -
 \frac{2}{3}[\p(\p^2-T)\ve^{(2)}_2]\ve^{(2)}_1) \frac{d}{dz}\right. +
\eq
$$
\left.
(\ve^{(2)}_2\p^2\ve^{(2)}_1-\ve^{(2)}_1\p^2\ve^{(2)}_2)\frac{d^2}{dz^2}\,,
\,\frac{2}{3}\oint\ve^{(2)}_1\p^5\ve^{(2)}_2\right)\,.
$$
Note, that the brackets (\ref{6.6}) are the Virasoro brackets and the whole
set of the commutation relations is their generalization on the second order
differential operators.

According with (\ref{dx}) the anchor action in ${\cal A}_3(D^*)$ has the form
\beq{6.9}
\de_{\ve^{(1)}}T=-2\p^3\ve^{(1)}+2T\p\ve^{(1)}+\p T\ve^{(1)}\,,
\eq
\beq{6.10}
\de_{\ve^{(1)}}W=-\p^4\ve^{(1)}+3W\p\ve^{(1)}+\p W\ve^{(1)}+T\p^2\ve^{(1)}\,,
\eq
\beq{6.11}
\de_{\ve^{(2)}}T=\p^4\ve^{(2)}-T\p^2\ve^{(2)}+(3W-2\p T)\p\ve^{(2)}+
(2\p W-\p^2T)\ve^{(2)}\,,
\eq
\beq{6.12}
\de_{\ve^{(2)}}W=\frac{2}{3}\p^5\ve^{(2)}-\frac{4}{3}T\p^3\ve^{(2)}-
2\p T\p^2\ve^{(2)}+
\eq
$$
(\frac{2}{3}T^2-2\p^2T+2\p W)\p\ve^{(2)}+
(\p^2W-\frac{2}{3}\p^3T+\frac{2}{3}T\p T)\ve^{(2)}\,.
$$
Thereby, we come to the Lie algebroid ${\cal A}_3(D^*)$ over $M_3(D^*)$.
This algebroid is nontrivial since
the structure functions in (\ref{6.8}) depend on  the projective connection $T$.

The SAJI (\ref{5.5}) in ${\cal A}_3(D^*)$ takes the form
\beq{6.14}
\lb\lb\ve^{(2)}_1,\ve^{(2)}_2\rb,\ve^{(2)}_3\rb^{(1)}-
(\ve^{(2)}_1\p\ve^{(2)}_2-\ve^{(2)}_2\p\ve^{(2)}_1)\de_{\ve^{(2)}_3}T+{\rm c.p.}(1,2,3)
=0,
\eq
\beq{6.15}
\lb\lb\ve^{(2)}_1,\ve^{(2)}_2\rb,\ve^{(1)}_3\rb^{(1)}-
(\ve^{(2)}_1\p\ve^{(2)}_2-\ve^{(2)}_2\p\ve^{(2)}_1)\de_{\ve^{(1)}_3}T=0.
\eq
The brackets here correspond to the product of
structure functions in the left hand side
of (\ref{5.5}) and the superscript $(1)$ corresponds to the ${\cal D}^1$ component.
For the  rest  brackets the Jacobi identity is the standard one.

The origin of the brackets and the anchor representations follow from the
matrix description of $\SLt$-opers (\ref{6.1}).
Consider the set $G_3(D^*)$ of automorphisms of the bundle $E_3$ over $D^*$
\beq{6.15a}
A\to f^{-1}\p f-f^{-1}Af\,,
\eq
that preserve the $\SLt$-oper structure
\beq{6.16}
f^{-1}\p f-
f^{-1}\thmat{0}{1}{0}{0}{0}{1}{W}{T}{0}f=
\thmat{0}{1}{0}{0}{0}{1}{W'}{T'}{0}\,.
\eq
It is clear that $G_3(D^*)$ is the Lie groupoid over  $M_3 (D^*)=\{W,T\}$ with
$l(f)=(W,T)$, $~r(f)=(W',T')$, $~f\to << W,T|f|W',T'>>$.
The left identity map is the $\SLt$  subgroup of $G_3$
$$
P\exp(-\int^z_{z_0} A(W,T))\cdot C\cdot P\exp(\int^z_{z_0} A(W,T))\,,
$$
where $C$ is an arbitrary matrix from $\SLt$ and $A(W,T))$ has the oper structure (\ref{6.1}).
The right identity map has the same
form with $(W,T)$ replaced by $(W',T')$.

The local version of (\ref{6.16}) takes the form
\beq{6.17}
\p X-\left[\thmat{0}{1}{0}{0}{0}{1}{W}{T}{0},X\right]=
\thmat{0}{0}{0}{0}{0}{0}{\de W}{\de T}{0}\,.
\eq
It is the sixth order linear differential system for the matrix elements of the
traceless matrix $X$. The matrix elements $x_{j,k}\in\Om^{(j-k,0)}(\Si_{g,n})$
 depend on two arbitrary fields $x_{23}=\ve^{(1)},~x_{13}=\ve^{(2)}$. The
solution takes the form
\beq{6.17b}
X=\thmat{x_{11}}{x_{12}}{\ve^{(2)}}{x_{21}}{x_{22}}
{\ve^{(1)}}{x_{31}}{x_{32}}{x_{33}}\,,
\eq
$$
x_{11}=\frac{2}{3}(\p^2-T)\ve^{(2)}-\p\ve^{(1)}\,,~
x_{12}=\ve^{(1)}-\p\ve^{(2)}\,,
$$
$$
x_{21}=\frac{2}{3}\p(\p^2-T)\ve^{(2)}-\p^2\ve^{(1)}+W\ve^{(2)}\,,~
x_{22}=-\frac{1}{3}(\p^2-T)\ve^{(2)}\,,
$$
$$
x_{31}=\frac{2}{3}\p^2(\p^2-T)\ve^{(2)}-\p^3\ve^{(1)}+\p(W\ve^{(2)})+W\ve^{(1)}\,,
$$
$$
x_{32}=\frac{1}{3}\p(\p^2-T)\ve^{(2)}-\p^2\ve^{(1)}+W\ve^{(2)}+T\ve^{(1)}\,,
$$
$$
x_{33}=-\frac{1}{3}(\p^2-T)\ve^{(2)}+\p\ve^{(1)}\,.
$$
The matrix elements of the commutator $[X_1,X_2]_{13}$, $[X_1,X_2]_{23}$
give rise to the brackets (\ref{6.6}),\\(\ref{6.7}), (\ref{6.8}). Simultaneously,
from (\ref{6.17}) one obtain the anchor action (\ref{6.9})--(\ref{6.12}).

\subsection{Lie algebroid over SL$(3,\mC)$-opers}

As in the previous examples, we can define the global algebroid ${\cal A}_3$ over
the space of opers $M_3$. The space of sections
${\cal G}_3\sim\G({\cal A}_3)=\{\ve^{(1)},~\ve^{(2)}\}$ are the
second order differential operators on $\Si_{g,n}$.
We assume that  $\ve^{(1)},~\ve^{(2)}$
vanish holomorphically at the marked points as
\beq{6.5}
\ve^{(1)}\sim r^{(1)}_a(z-x_a)+o(z-x_a)\,,~~
\ve^{(2)}\sim r^{(2)}_a(z-x_a)^2+o(z-x_a)^2\,~~r^{(j)}\neq 0\,.
\eq
Note that these asymptotics are consistent with the Lie brackets and with asymptotics of
$T$ (\ref{6.3}) and $W$ (\ref{6.4}).

Consider the cohomology of ${\cal A}_3$.
There exists a nontrivial cocycle corresponding to $H^1({\cal A}_3)$
with two components
\beq{6.17a}
c^{(1)}=\int_{\Si_{g,n}}\ve^{(1)}\bp T\,,
~~c^{(2)}=\int_{\Si_{g,n}}\ve^{(2)}\bp W\,.
\eq
It follows from (\ref{6.3}), (\ref{6.4}) and (\ref{6.5}) that the contributions
 from the marked points are equal
$$
c^{(1)}\rar \sum_{a=1}^n r^{(1)}_aT_{-2,a}\,,~~
c^{(2)}\rar \sum_{a=1}^n r^{(2)}_aW_{-3,a}\,.
$$

The cocycle leads to the shift of the anchor action
$$
\hat{\de}_{\ve^{(j)}}f(W,T)=\lan \de_{\ve^{(j)}}W|\frac{\de f}{\de W}\ran   +
\lan \de_{\ve^{(j)}}T|\frac{\de f}{\de T}\ran   +c^{(j)}\,.
$$

There exists the $2g$ central extensions $c_\al$ of  ${\cal G}_3$,
provided by the nontrivial cocycles from  $H^2({\cal A}_3,M_3)$. They are the
non-contractible contour integrals $\ga_\al$
\beq{6.16a}
c_\al(\ve^{(j)}_1,\ve^{(k)}_2)=\oint_{\ga_\al}\la(\ve^{(j)}_1,\ve^{(k)}_2),
~~(j,k=1,2)\,,
\eq
where
$$
\la(\ve^{(1)}_1,\ve^{(1)}_2)=-2\ve_1^{(1)}\p^3\ve_2^{('1)}\,,~~
\la(\ve^{(1)}_1,\ve^{(2)}_2)=\ve^{(1)}_1\p^4\ve^{(2)}_2\,,
$$
$$
\la(\ve^{(2)}_1,\ve^{(2)}_2)=\frac{2}{3}
\ve_1^{(2)}\p^5\ve_2^{(2)}\,.
$$
It can be proved that $sc_\al=0$
(\ref{5.9}) and that $c_\al$ are not exact. The proof is based on the matrix realization
of $\G({\cal A}_3)$ (\ref{6.17b}) and the two-cocycle (\ref{8.3}) of ${\cal A}_{SL_N}$.
These cocycles allow us to construct the extended brackets:
$$
\lb(\ve_1^{(j)},\sum_\al k^{(j)}_\al),(\ve_2^{(m)},\sum_\al k^{(m)}_\al)\rb_{c.e.}=
(\lb\ve_1^{(j)},\ve_2^{(m)}\rb,\sum_\al c_\al(\ve_1^{(j)},\ve_2^{(m)}))\,.
$$

\subsection{Hamiltonian algebroid over $W_3$-gravity}

Let ${\cal R}_3=Aff\, T^*M_3$ be the affinization over the cotangent bundle $T^*M_3$
 to the space of $\SLt$-opers $M_3$. The dual fields are the Beltrami differentials
$\mu$ and the differentials $\rho\in\Om^{(-2,1)}(\Si_{g,n})$.
We assume that   near the marked points $\rho$
 has the form
\beq{6.19}
\rho|_{z\rar x_a}\sim
(t^{(2)}_{a,0}+t^{(2)}_{a,1}(z-x^0_a))\bp\chi_a(z,\bz)\,,
\eq
and $\mu$ as before satisfies (\ref{mu}).
The space   ${\cal R}_3$ is the classical phase space for the $W_3$-gravity
\cite{P,BFK,GLM}. The symplectic form on ${\cal R}_3$ has the canonical form
$$
\om=\int_{\Si_{g,n}}D T\wedge D\mu+D W\wedge D\rho\,.
$$

According to the general theory the anchor (\ref{6.9})--(\ref{6.12})
can be lifted from $M_3$ to ${\cal R}_3$.
This lift is nontrivial owing to the cocycle (\ref{6.17a}).
It follows from (\ref{5.16a}) that the anchor action on $\mu$ and
$\rho$ takes the form
\beq{6.21}
\de_{\ve^{(1)}}\mu=-\bar{\p}\ve^{(1)}-\mu\p\ve^{(1)}+\p\mu\ve^{(1)}-
\rho\p^2\ve^{(1)}\,,
\eq
\beq{6.22}
\de_{\ve^{(1)}}\rho=-2\rho\p\ve^{(1)}+\p\rho\ve^{(1)}\,,
\eq
\beq{6.23}
\de_{\ve^{(2)}}\mu=\p^2\mu\ve^{(2)}-\frac{2}{3}\left[(\p(\p^2-T)\rho)\ve^{(2)}
-(\p(\p^2-T)\ve^{(2)})\rho\right]\,,
\eq
\beq{6.24}
\de_{\ve^{(2)}}\rho=-\bar{\p}\ve^{(2)}+(\rho\p^2\ve^{(2)}-\p^2\rho\ve^{(2)})
+2\p\mu\ve^{(2)}-\mu\p\ve^{(2)}\,.
\eq
There are two Hamiltonians, defining by the anchor
$$
h^{(1)}=\int_{\Si_{g,n}}(\mu\de_{\ve^{(1)}}T+\rho\de_{\ve^{(1)}}W)+c^{(1)}\,,~~
h^{(2)}=\int_{\Si_{g,n}}(\mu\de_{\ve^{(2)}}T+\rho\de_{\ve^{(2)}}W)+c^{(2)}\,.
$$
After the integration by parts they take the form
$$
h^{(1)}=\int_{\Si_{g,n}}\ve^{(1)}F^{(1)}\,,~~
h^{(2)}=\int_{\Si_{g,n}}\ve^{(2)}F^{(2)}\,,
$$
where $F^{(1)}\in\Om^{(2,1)}(\Si_{g,n})$, $F^2\in\Om^{(3,1)}(\Si_{g,n})$
\beq{F1}
F^{(1)}=-\bp T-\p^4\rho+T\p^2\rho-(3W-2\p T)\p\rho-
\eq
$$
-(2\p W-\p^2 T)\rho+2\p^3\mu-2\p T\mu-\p T\mu\,,
$$
\beq{F2}
F^{(2)}=-\bp W-\frac{2}{5}\p^5\rho+\frac{4}{3}T\p^3\rho+2\p T\p^2\rho+
(-\frac{2}{3}T^2+2\p^2 T-2\p W)\p\rho+
\eq
$$
+(-\p^2W+\frac{2}{3}\p^3T-\frac{2}{3}T\p T)\rho+\p^4\mu
-3W\p\mu-\p W\mu-T\p^2\mu\,.
$$
The Hamiltonians carry out the moment map
$$
m=(m^{(1)}=F^{(1)},m^{(2)}=F^{(2)}): {\cal R}_3\to {\cal G}^*_3\,.
$$
The elements of the dual space ${\cal G}^*_3$ are singular at the marked points. In addition
to $y$ (\ref{8.13a}) there are $v$ dual to $\ve^{(2)}$ (\ref{6.5})
$$
v\sim c_{1,a}\p^2\de(x_a)+c_{2,a}\p^3\de(x_a)+\ldots\,.
$$
Let $m^{(1)}$ is defined as in (\ref{8.15}) and
$$
m^{(2)}=\sum_{a=1}^nW^a_{-3}\p^2\de(x_a)\,.
$$
Then the coadjoint action of $G_2$ preserve $m=(m^{(1)},m^{(2)})$.
The moduli space  ${\cal W}_3$ of the $W_3$-gravity
( ${\cal W}_3$-geometry) is the symplectic quotient with respect to
the groupoid
 action
$$
{\cal W}_3={\cal R}_3// G_2=\{F^1=m^{(1)},F^2=m^{(2)}\}/G_2.
$$
It has dimension is
$\dim{\cal W}_3=16(g-1)+6n$.
The term $6n$ comes from the coefficients
$T_{-1}^a,W^a_{-1},W^a_{-2}$, and the dual to them
$t^{(1)}_{a,0},t^{(2)}_{a,0},t^{(2)}_{a,1}$,  $(a=1,\dots,n)$ in (\ref{mu}) and (\ref{6.19}).

The moment equations $F^{(1)}=m^{(1)},~F^{(2)}=m^{(2)}$ are the consistency conditions
for the linear system
\beq{6.26a}
\left\{
\begin{array}{l}
(\p^3-T\p-W)\psi(z,\bz)=0\,,\\
\left(\bp +(\mu-\p\rho)\p +\rho\p^2
+\frac{2}{3}(\p^2-T)\rho-\p\mu
\right)\psi(z,\bz)=0\,,
\end{array}
\right.
\eq
where $\psi(z,\bz)\in\Om^{-1,0}(\Si_{g,n})$. We will prove this statement below.
The last equation  represents the deformation of the antiholomorphic
operator $\bp$ (or more general $\bp+\mu\p$ as in (\ref{8.16}))
by the second order differential
operator $\p^2$. The left hand side is the exact form of the deformed operator
when it acts on $\Om^{-1,0}(\Si_{g,n})$. This deformation cannot
be supported by the structure of a Lie algebra  and one leaves with
the algebroid symmetries.

The prequantization of ${\cal W}_3$ can be realized in the space of sections
of a linear bundle ${\cal L}$ over the space of orbits
 $\ti{M}_3\sim M_3/G_3$. The sections are functionals
$\Psi(T,W)$ on $M_3$ such that
$\hat{\delta}_{\ve^{(j)}}\Psi(T,W)=0,~(j=1,2)$.
The bundle ${\cal L}$ can be identified with the determinant bundle
 $\det(\p^3-T\p-W)$.

Instead of the symplectic reduction one can apply the BRST construction.
The cohomology of the moduli space  ${\cal W}_3$ are
isomorphic to
$H^j(Q)$. To construct the BRST complex we introduce the ghosts
fields $\eta^{(1)},\eta^{(2)}$ and their momenta ${\cal P}^{(1)},{\cal P}^{(2)}$.
Then it follows from Theorem 2.1 that for
$$
\Om=\sum_{j=1,2}h^{(j)}(\eta^{(j)})+
\oh\sum_{j,k,l=1,2}\int_{\Si_{g,n}}(\lb\eta^{(j)},\eta^{(k)}\rb{\cal P}^{(l)})
$$
the operator  $QF=\{F,\Om\}$ is nilpotent and define the BRST cohomology
in the complex
$$
\bigwedge{}^\bullet({\cal G}_3\oplus{\cal G}_3^*)\otimes C^\infty({\cal R}_3)\,.
$$
\bigskip

\subsection{Chern-Simons derivation}

We follow here the derivation of $W$-gravity proposed in Ref.\cite{BFK}.
We only add in the construction a contribution of the Wilson lines
 due to the presence of the marked points on $\Si_{g,n}$.

Consider the Chern-Simons functional on $\Si_{g,n}\oplus {\mR}^+$
$$
S=\int_{\Si_{g,n}\oplus \mR^+}\tr({\bf A}d{\bf A}+\frac{2}{3}{\bf A}^3)+
\sum_{a=1}^n\int_{\mR^+}\tr(p_ag_a^{-1}\p_tg_a)\,,~~
({\bf A}=(A,\bA,A_t))\,.
$$
 Introduce $n$ Wilson lines $W_a(A_t)$
along the time directions and located at the marked points
$$
W_a(A_t)=P\exp \tr(p_a\int A_t),~a=1,\ldots,n\,.
$$
In the hamiltonian picture the components
$A,\bA,\bfp=(p_1,\ldots,p_n),\bfg=(g_1,\ldots,g_n)$ are elements of the phase space
 with the symplectic form
(\ref{8.4a}) while $A_t$ is the Lagrange multiplier for the first class constraints
(\ref{8.5a}).

The phase space ${\cal R}_3$ can be derived from the Chern-Simons phase space.
 The flatness condition (\ref{8.5a}) generates
the gauge transformations
\beq{5.31}
A\to f^{-1}\p f-f^{-1}Af\,,~~\bA\to f^{-1}\bp f-f^{-1}\bA f\,,~~
p_a\to f_a^{-1}p_af_a,~~g_a\to g_af_a\,.
\eq
The result of the gauge fixing with respect to the whole gauge group $G_{\SLt}$
is the moduli
space ${\cal M}^{flat}_3$ of the flat $\SLt$ bundles over $\Si_{g,n}$.

Let $P$ be the maximal parabolic subgroup of $\SLt$ of the form
$$
P=\thmat{*}{*}{0}{*}{*}{0}{*}{*}{*}\,,
$$
and $G_P$ be the corresponding gauge group.
First, we partly fix the gauge with respect to $G_P$.
A generic connection $\nabla$ can be gauge transformed by $f\in G_P$ to
  the form (\ref{6.1}).
It follows from (\ref{8.5a}) that $A$ has simple poles at the marked points.
To come to $M_3$
one should respect the behavior of the matrix elements at the marked points
(\ref{6.3}), (\ref{6.4}). For this purpose we use an additional singular gauge transform
by the diagonal matrix
$$
h=\prod_{a=1}^n\chi_a(z,\bz)\di(z-x_a,1,(z-x_a)^{-1})\,.
$$
The resulting gauge group we denote $G_{(P,h)}$.

The form of $\bA$ can be read off from (\ref{8.5a})
\beq{5.32}
\bA=\thmat{a_{11}}
{a_{12}}
{-\rho}
{a_{21}}
{a_{22}}
{-\mu}
{a_{31}}
{a_{32}}
{a_{33}}\,,
\eq
$$
a_{11}=-\frac{2}{3}(\p^2-T)\rho+\p\mu\,,~~
a_{12}=-\mu+\p\rho\,,
$$
$$
a_{21}=-\frac{2}{3}\p(\p^2-T)\rho+\p^2\mu-W\rho\,,~~
a_{22}=\frac{1}{3}(\p^2-T)\rho\,,
$$
$$
a_{31}=-\frac{2}{3}\p^2(\p^2-T)\rho+\p^3\mu-\p(W\rho)-W\mu\,,
$$
$$
a_{32}=-\frac{1}{3}\p(\p^2-T)\rho+\p^2\mu-W\rho-T\mu\,,~~
a_{33}=\frac{1}{3}(\p^2-T)\rho-\p\mu\,.
$$
The flatness (\ref{8.5a})  for the special choice $A$ (\ref{6.1}) and
$\bA$ (\ref{5.32}) gives rise to
the moment constraints $F^{(2)}=0,~F^{(1)}=0$. Namely, one has
$F(A,\bA)|_{(3,1)}=F^{(2)}$ (\ref{F1}), $F(A,\bA)|_{(2,1)}=F^{(1)}$ (\ref{F2}),
while the other matrix
elements of $F(A,\bA)$ vanish identically.
At the same time, the matrix linear
system (\ref{8.5c}) coincides with  (\ref{6.26a}).
In this way, we come to the matrix description of the moduli space
${\cal W}_3$.

The cocycles $c_\al(\ve^{(j)}_1,\ve^{(k)}_2)$ (\ref{6.16a}) can be derived from
the two-cocycle (\ref{8.3}) of ${\cal A}_{SL_3}$. Substituting in (\ref{8.3})
the matrix realization of $\G({\cal A}_3)$ (\ref{6.17b}), one come to (\ref{6.16a}).

The groupoid action on $A,\bA$ plays the role of the rest gauge
transformations that complete the $G_P$ action to the $G_{SL_3}$ action.
The algebroid symmetry arises in this theory as a result of the partial
gauge fixing by $G_{(P,h)}$. Thus we come to the following diagram

\bigskip
$$
\begin{array}{rcccl}
           &\fbox{${\cal R}_{SL_3}$}&                       &                    &\\
           &       |                      &\searrow{G_{(P,h)}}&           &\\
G_{\SLt} &        |                     &                    &\fbox{${\cal R}_3$}&\\
           &\downarrow    &                   &\downarrow &G_3\\
           &\fbox{${\cal M}^{flat}_{SL_3}$}&       &\fbox{${\cal W}_3$}&\\
\end{array}
$$
\bigskip

The tangent space to ${\cal M}^{flat}_{SL_3}$ at the point $A=0,\bA=0,p_a=0,g_a=id$
coincides with the tangent space to ${\cal W}_3$ at the point
$W=0,T=0,\mu=0,\rho=0$. Their dimension is $16(g-1)+6n$. But their global
structure is different and the diagram cannot be closed by the horizontal
isomorphisms. The interrelations between ${\cal M}^{flat}_{SL_3}$ and ${\cal W}_3$
were analyzed in \cite{H2,Go}.

\section{${\cal W}_N$-gravity and general deformations of complex structures}
\setcounter{equation}{0}

In this section we present the general deformation of complex
structures by the Volterra operators. To construct the Lie algebroid over
 the space of $\GL$-opers we use another form of the pairing. Instead of the
differential operators and the pairing (\ref{pa}) we consider the Volterra operators
 that come from the pairing (\ref{pa1}).
We start with the description of the local AGD algebroid following Ref.\,
\cite{GD} and then give its global version. The passage from the Lie
algebroid to the corresponding Hamiltonian algebroid allows us to describe
the deformations of the complex structures.

\subsection{Local AGD algebroid }

Consider the set $B=\Psi DO(D^*)$ of pseudo-differential symbols on the disk $D^*$.
It is a ring of formal Laurent series
$$
B=B((\p^{-1}))=
\{B_{r,N}(D^*)\,,~~r,N\in\mZ\}=\{X(z,\p)\}\,,
$$
\beq{9.1}
X(z,\p)=\sum_{k=-\infty}^ra_k(z)\p^k,~~
( a_k(z)\in \Om^{-N-k}(D^*) \,.
\eq
 The multiplication on $B$ is defined as the non-commutative
multiplication of their symbols
\beq{pr}
X(z,\la)\circ Y(z,\la)=
\sum_{k\geq 0}\f1{k!}\frac{\p^k}{\p\la^k} X(z,\la)\frac{\p^k}{\p z^k}Y(z,\la)\,.
\eq
In what follows we drop the multiplication symbol $\circ$.

Note that $B_{r,N}(D^*)\in\Psi DO(D^*)$ can be considered as the formal map
of the sheaves
\beq{B}
B_{r,N}(D^*)~:~\Om^{\frac{N+1}{2}}\to\Om^{-\frac{N-1}{2}}\,.
\eq
Let $L_N=\p^N+W_1\p^{N-1}+\ldots+W_N$ be $\GL$-oper on $D^*$. Then we have
$$
L_NX~:~\Om^{-\frac{N-1}{2}}\to\Om^{-\frac{N-1}{2}}\,,~~~
L_NX~:~\Om^{\frac{N+1}{2}}\to\Om^{\frac{N+1}{2}}\,,
$$
where the product is defined as (\ref{pr}). There is the functional
on $\Om^{-\frac{N-1}{2}}$
\beq{pa1}
\lan L_NX\ran=\f1{2\pi}\oint_{|z|=1} Res (L_NX)dz\,,
\eq
where $Res=a_{-1}$. The important property is that
$\lan L_NX\ran=\lan XL_N\ran$.

The AGD brackets on the space $M^G_N(D^*)$ of $\GL$-opers
$L_N$ are defined as follows.
The space of section of $T^*M^G_N(D^*)$ can be identified
with  the quotient space of the Volterra operators
\beq{9.1a}
\G(T^*M^G_N(D^*))= B_{0,N}(D^*)/B_{-N-2,N}(D^*)
\eq
For $L_N $ and $X\in \G(T^*M^G_N(D^*))$ define the functional
$l_X=\lan L_NX\ran$.
In particular, $W_j(z)=\lan L_N\de(z)\p^{-j-1}\ran$.
The AGD brackets have the form
\beq{9.2}
\{l_X,l_Y\}=\lan L_NX(L_NY)_+-XL_N(YL_N)_+\ran\,,
\eq
where $X_+=\sum_{k=0}^Na_k\p^k $ is the differential part of $X$.

Using the general prescription we define the Lie brackets
(\ref{2.3}) in the space  of sections of $ T^*M^G_N(D^*)$
\beq{9.3}
\lb X,Y\rb=X(L_NY)_+-(YL_N)_+X+(XL_N)_+Y-Y(L_NX)_+\,.
\eq
and the anchor map
\beq{9.4}
\de_YL_N=(L_NY)_+L_N-L_N(YL_N)_+\,.
\eq
We can rewrite (\ref{9.2}) in the form of the "Poisson-Lie brackets"
$$
\{l_X,l_Y\}=\oh\lan \lb X,Y\rb L_N\ran\,.
$$
The coefficient $1/2$ arises from the quadratic form of the Poisson bivector.

In this way we have constructed the local Lie algebroid ${\cal A}^G_N(D^*)$
over the space of the local $\GL$ opers $M^G_N(D^*)$.

To be the $\SLN$-oper $L_N=\sum_{k=0}^Na_k\p^k$ should satisfy the second
class constraints
$$
W_1(z)=0\,,~~~W_{1}=\lan L_N,\de(z)\p^{-N}\ran \,.
$$
since $\{W_{1}(z), W_{1}(w)\}\sim\de(z-w)$. The functional $ W_{1}(z)$
generates the vector field on $M^G_N(D^*)$
$$
\de_Y L_N=\{W_{1},L_N\}=\oh\lan\lb\de(z)\p^{-N},Y\rb L_N\ran=
\lan\de(z) \p^{-N}\de_YL_N\ran\,.
$$
If $\de_Y L_N=0$ the transformed oper is the $\SLN$-oper.
It means that $ Res [Y,L_N]=0$.
In this way the brackets (\ref{9.2}) is the generalization of (\ref{vir}) $(N=2)$
and (\ref{6.9a})-(\ref{6.12a}) $(N=3)$ on arbitrary $N$.
The relations between the Poisson manifolds $M^G_N(D^*)$ and $M_N(D^*)$
were discussed also in Ref.\,\cite{KZ}.

\subsection{Poisson sigma-model}

Here we construct the Poisson sigma-model with the target space $M^G_N(D^*)$
following {\bf 3.2}.
We modify the notion of $\Psi DO(D^*)$ assuming that instead of (\ref{B})
we have
\beq{B1}
B_{(r,N,1)}(D^*)= B_{r,N}(D^*)\otimes \bar{K}(D^*)\,,~~
B_{(r,N,1)}(D^*)\,:
\,\Om^{(\frac{N+1}{2},1)}\to\Om^{(-\frac{N-1}{2},1)}\,.
\eq
The affinization ${\cal R}_N=Aff\,T^*{\cal A}_N(D^*)$ of $T^*M^G_N(D^*)$
is defined by the pair of fields $(L_N,\xi)$,
 where $\xi\in B_{(0,N,1)}/ B_{(-N-2,N,1)}$
is the dual field to an oper with respect to the integral over $D^*$.
The space ${\cal R}_N $ is the phase space of the Poisson sigma model
with the canonical symplectic form
\beq{9.6}
\om=\int_{D^*}Res (DL_N \wedge D\xi) \,.
\eq

The one-cocycle $c(L,Y)\in H^1({\cal A}_N(D^*))$
\beq{9.5}
c(L,Y)=\int_{D^*}Res(Y\bp L_N)\,,~~(Y\in B_{0,N})
\eq
provides the lift of the anchor action (\ref{9.4}) on ${\cal R}_N $
\beq{9.7}
\de_Y\xi=-\bp Y+
Y(L_N\xi)_+-(\xi L_N)_+Y+(YL_N)_+\xi-\xi(L_NY)_+\,.
\eq
Along with (\ref{9.4}) this action defines the canonical transformations
of (\ref{9.6}). They are generated by the Hamiltonian
\beq{9.8}
h_Y= \int_{D^*} Res(\xi\de_YL_N)+ c(L,Y)=
\int_{D^*} Res\left (Y (
\bp L_N-(L_N\xi)_+L_N+L_N(\xi L_N)_+)
\right)\,.
\eq
This expression yields  the anchor map of the Hamiltonian
algebroid ${\cal A}_N^H(D^*)$
related to ${\cal A}_N(D^*)$. The canonical transformations by $B_{0,N}(D^*)$
$$
\de_YL_N=\{h_Y,L_N\}\,,~~\de_Y\xi=\{h_Y,\xi\}\,,~~(Y\in B_{0,N}(D^*))
$$
 corresponding to the form (\ref{9.6}) are generated by
 the first class constraints
\beq{9.9}
F:=\bp L_N-(L_N\xi)_+L_N+L_N(\xi L_N)_+=0\,.
\eq

Define the operator
\beq{9.11}
A=\bp-(L_N\xi)_+\,:\,
\Om^{(\frac{N+1}{2},0) }(D^*)\to\Om^{(\frac{N+1}{2},1)} (D^*)
\eq
and the dual operator
$$
A^*=\bp+(\xi L_N)_+ \,:\,
 \Om^{(-\frac{N-1}{2},0)} (D^*)\to\Om^{(-\frac{N-1}{2},1)} (D^*)\,.
$$
Let $\psi=(\psi^-,\psi^+)$, $\psi^-\in\Om^{(-\frac{N-1}{2},0)} (D^*)$,
$ \psi^+\in\Om^{(\frac{N+1}{2},0)} (D^*)$.
 Following Lemma 3.2 we conclude that the quadratic constraints
(\ref{9.9}) are equivalent to the linear problem
\beq{9.10}
\left\{
\begin{array}{l}
L_N\psi^-(z,\bz)=0\,,\\
(\bp-(L_N\xi)_+)\psi^-(z,\bz)=0\,.
\end{array}
\right.
\eq
The second equation can be replaced on
\beq{9.12}
(\bp+(\xi L_N)_+)\psi^+(z,\bz)=0\,.
\eq
In this way the local oper $L_N$ along with the dual element $\xi$
defines {\it the deformation of the complex structure} on the disk $D^*$.
The second equation (\ref{9.10}) is equivalent to the deformed
holomorphity condition for the sections $\Om^{(-\frac{N-1}{2},0)}$.

Let $G^H_N(D^*)$ be the Hamiltonian Lie groupoid corresponding to
${\cal A}_N^H(D^*)$.
The reduced phase space is the symplectic quotient
$$
{\cal R}^{red}_N={\cal R}//G^H_N(D^*)=\{F=0\}/G^H_N(D^*)\,.
$$

\subsection{Global algebroid}

As we have noted in previous Section the $\SLN$-opers are well defined
on the curve $\Si_{g,n}$.
The global algebroid ${\cal A}_N$ over the space of $\GL$-opers $M^G_N$
on $\Si_{g,n}$ is independent on the choice of $D^*$.
The space of its sections ${\cal G}_{N}\sim\G({\cal A}_N)$ is the
space of the Volterra operators on $\Si_{g,n}$ with smooth
coefficients (see(\ref{9.1a})).
Near a marked point $x_a$ with a local coordinate $z$ for $X\in {\cal G}_{N}$
we have
$$
X=\ep^{(1)}\p_z^{-1}+\ep^{(2)}\p_z^{-2}+\ldots+ \ep^{(N)}\p_z^{-N}+
\ep^{(N+1)}\p_z^{-N-1}\,,
$$
where
$$
\ve^{(j)} \sim r_a ^{(j)}(z-x_a)^j+o(z-x_a)^j\,,~~ r_a ^{(j)}\neq 0\,.
$$
The anchor action of $X$ is defined as before by (\ref{9.4}).

The one-cocycle representing $H^1({\cal A}_N)$ comes from
 the integration over $\Si_{g,n}$
$$
c=\int_{\Si_{g,n}}X{\bp}L_N\,.
$$

Let
$$
B_{(r,N,1)}( \Si_{g,n})= B_{r,N}(\Si_{g,n})\otimes \bar{K}(\Si_{g,n})\,,
$$
$$
B_{(r,N,1)}( \Si_{g,n})\,:
\,\G(\Om^{(\frac{N+1}{2},1)} ( \Si_{g,n}))
\to\G(\Om^{(\frac{-N-1}{2},1)} ( \Si_{g,n}))\,.
$$
The affinization ${\cal R}_N=Aff\,T^*{\cal A}_N( \Si_{g,n})$
of $T^*M_N( \Si_{g,n})$
is defined by the fields $\xi$ from the quotient
$\xi\in B_{(0,N,1)}( \Si_{g,n})/ B_{(-N-2,N,1)}( \Si_{g,n})$.
Near the marked points
$$
\xi=\sum_{j=1}^{N+1}\nu_j\p^{-j},~~(\nu_{N+1}=1)\,,
$$
$$
\nu_j|_{z\to x_a}=
(t^{(j)}_{a,0}+\ldots+ t^{(j)}_{a,j-1}(z-x_a)^{j-1})\bar{\p}\chi_a(z,\bz)\,,
~~(\nu_j\in \Om^{(j-N,1)}(\Si_{g,n}))\,.
$$
The symplectic form on  ${\cal R}_N$ is
$$
\om= \int_{\Si_{g,n}}Res\,(DL_N \wedge D\xi)\,.
$$
The anchor action on $M^G_N$ (\ref{9.4}) can be lifted from $M^G_N$
to ${\cal R}_N$ as canonical transformations of $\om$.   They are generated by the Hamiltonians
$$
h_Y=\int_{\Si_{g,n}}Res\,(\xi\de_{Y}L_N)+c(Y,L_N)\,.
$$
In addition to (\ref{9.4}) we have the action on the dual variables are
\beq{9.4a}
\de_Y\xi=\{ h_Y,\xi\}\,.
\eq
The Hamiltonians can be represented in the form
$$
h_Y=\int_{\Si_{g,n}}Res\,(YF(L_N,\xi))\,,
$$
where $F(L_N,\xi)$ is defined by (\ref{9.9}).
They lead to the moment map
$$
m:{\cal R}_N\to {\cal G}_{N}^*\,,~~
m= F(L_N,\xi)\,.
$$
We take
$$
m=\sum_{a=1}^n \sum_{j=1}^NW_{-j}^a(j)\p^{j-1}|_{z=x_a} \,,
$$
where $ W_{-j}^a(k)$ are defined in (\ref{6.4a}). The moment equation
$F= m$
 is equivalent to the linear problem (\ref{9.10}), (\ref{9.12}) on $\Si_{g,n}$.
In this way we come to the deformed $\bp$ operator
$\bp-(L_N\xi)_+$ $(\bp+(\xi L_N)_+)$ acting on the space
$\Om^{(-\frac{N-1}{2},0)} (\Si_{g,n})$ $(\Om^{(\frac{N+1}{2},0)} (\Si_{g,n}))$.

The moduli space of the generalized
complex structures is a part of the symplectic quotient
${\cal W}_N\sim{\cal R}_N//G_{N}$,
where $G_N$ is the corresponding Hamiltonian groupoid. The cohomology of
${\cal W}_N $ are defined by the BRST operator
$$
\Om=h_\eta+
\oh\int_{\Si_{g,n}}Res\left(
(\lb\eta,\eta'\rb{\cal P}
\right)\,,
$$
where $\eta$ is the ghost field corresponding to the gauge field $Y$
and ${\cal P}$ is its momenta.

\section{Concluding Remarks}
 Let summarize the results and discuss some open problems.

(i)
The Hamiltonian algebroid is a bundle over a Poisson manifold with a
Lie algebra structure on its sections and the anchor map to the Hamiltonian
vector fields.
 The special kind of the Hamiltonian algebroids are defined over
affine space of
cotangent bundles. They are lift of the Lie algebroids defined over
the base of the cotangent bundles.
The lifts are classified by the first cohomology of the Lie algebroids.
The Hamiltonian algebroids  of this type are most closed to the Lie algebras
of Hamiltonian vector fields and has the same structure of the BRST operator.

(ii) The Lie algebroid over the space of $\SLt$-opers on a Riemann curve with
marked points has the space of the second order differential operators as
the space of its sections. It contains the Lie subalgebra of the first order
differential operators. After change the behavior of their coefficients
at the marked points this subalgebra coincides with the Krichever-Novikov
algebra \cite{KN}. It will be interesting to lift this correspondence to the
second order differential operators. Another open question is the structure of
opers and Lie algebroids defined on Riemann curves with double marked points.

(iii) In the limit $N\to\infty$ the structure
of the strongly homotopy Lie algebras
 should be recovered. In our approach this limit looks obscure.

(iv) The Chern-Simons derivation of the Hamiltonian algebroid in $W_N$-gravity
explain the origin of the algebroid symmetry as a result of the two step gauge
fixing. It will be plausible to have the same universal construction for
an arbitrary Poisson sigma-model.

\section*{Acknowledgments}
\addcontentsline{toc}{section}{\numberline{}Acknowledgments}
 This work was originated from discussions with S. Barannikov
concerning generalized deformations of complex structures during
the visit of the second author the IHES (Bur-sur-Yvette) in 1999.
We benefited also from valuable discussions with  A. Gerasimov, S. Liakhovich and M. Grigoriev.
The work of A.L. is supported in part by grants RFFI-98-01-00344
and 96-15-96455 for support of scientific schools.
The work of M.O. is supported in part by grants RFFI-00-02-16530,
 INTAS 99-01782, DFG-RFBR project 436 RUS 113/669
   and 96-15-96455 for support of scientific schools.

\addcontentsline{toc}{section}{\numberline{}References}
\setcounter{equation}{0}

\small{

}
\end{document}